\documentclass[aps,twocolumn,floatix,amssymb,showpacs,amsmath, superscriptaddress,longbibliography]{revtex4-1}
\usepackage{graphicx}
\usepackage{epsfig}
\usepackage[table,dvipsnames]{xcolor}
\usepackage{amsmath,amssymb,amsthm}
\usepackage[utf8]{inputenc}
\usepackage[colorlinks, urlcolor=blue, linkcolor=red]{hyperref}

\usepackage{soul}

\newcommand{\beqa}{\begin{eqnarray}}
\newcommand{\eeqa}{\end{eqnarray}}

\DeclareMathOperator{\Tr}{Tr}
\DeclareMathOperator{\Det}{Det}
\newcommand{\vect}[1]{\ensuremath{\mathbf{#1}}}

\renewcommand{\boxed}[2]{\textcolor{#1}{%
\tikz[baseline={([yshift=-1ex]current bounding box.center)}] \node [rectangle, minimum width=1ex,rounded corners,draw] {\normalcolor\m@th$\displaystyle#2$};}}

\newcounter{appsection}
\newcounter{appsubsection}[appsection]

\usepackage{bookmark}

\begin{document}

\title{Optimality and Noise-Resilience of Critical Quantum Sensing}
\author{U. Alushi}
\thanks{These two authors contributed equally}
\affiliation{Department of Information and Communications Engineering, Aalto University, Espoo, 02150 Finland}
\affiliation{Institute for Complex Systems, National Research Council (ISC-CNR), Via dei Taurini 19, 00185 Rome, Italy}
\author{W. G{\'o}recki}
\thanks{These two authors contributed equally}
\affiliation{INFN Sez.~Pavia, via Bassi 6, I-27100 Pavia, Italy}
\author{S. Felicetti}
\email{felicetti.simone@gmail.com}
\affiliation{Institute for Complex Systems, National Research Council (ISC-CNR), Via dei Taurini 19, 00185 Rome, Italy}
\affiliation{Physics Department, Sapienza University, P.le A. Moro 2, 00185 Rome, Italy}
\author{R. Di Candia}
\email{rob.dicandia@gmail.com}
\affiliation{Department of Information and Communications Engineering, Aalto University, Espoo, 02150 Finland}
\affiliation{Dipartimento di Fisica, Universit\`a degli Studi di Pavia, Via Agostino Bassi 6, I-27100, Pavia, Italy}

\begin{abstract}
We compare critical quantum sensing to passive quantum strategies to perform frequency estimation, in the case of single-mode quadratic Hamiltonians.  We show that, while in the unitary case both strategies achieve precision scaling quadratic with the number of photons, in the presence of dissipation this is true only for critical strategies. We also establish that working at the exceptional point or beyond threshold provides sub-optimal performance. This critical enhancement is due to the emergence of a transient regime in the open critical dynamics, and is invariant to temperature changes. When considering both time and system size as resources, for both strategies the precision scales linearly with the product of the total time and the number of photons, in accordance with fundamental bounds.
However, we show that critical protocols outperform optimal passive strategies if preparation and measurement times are not negligible. Our results are applicable to a broad variety of critical sensors whose phenomenology can be reduced to that of a single-mode quadratic Hamiltonian, including systems described by finite-component and fully-connected models.
\end{abstract}

\maketitle
{\it Introduction.---} The susceptibility developed in proximity of critical phase transitions (PTs) is a valuable resource in metrological tasks. This concept is widely exploited in advanced sensors such as transition-edge detectors and bubble chambers. However, these devices make use of a classical sensing strategy, and they are not optimal from a quantum-metrology perspective~\cite{Paris2009,RevModPhys_QSensing}. The recently introduced research field of critical quantum sensing (CQS) consists of leveraging quantum PTs to design quantum-enhanced sensors~\cite{Zanardi2008,ivanov_adiabatic_2013,Bina2016,Lorenzo2017,Fre18,Ivanov2020,invernizzi2008Optimal,Mirkhalaf2020,Niezgoda2021,Tsang2013,macieszczak_dynamical_2016,Cabot24}. In the last few years, it has been theoretically shown that it is possible to achieve quantum advantage in sensing exploiting both static~\cite{Zanardi2008,ivanov_adiabatic_2013,Bina2016,Lorenzo2017,Fre18,Ivanov2020,invernizzi2008Optimal,Mirkhalaf2020,Niezgoda2021} and dynamical~\cite{Tsang2013,macieszczak_dynamical_2016,Cabot24} critical properties of many-body quantum systems. First experimental demonstrations of quantum-enhanced sensing have been achieved with Rydberg atoms~\cite{Ding2022} and nuclear magnetic resonance techniques~\cite{Liu2021}.

\begin{figure}[t!]
  \includegraphics[width=0.48
\textwidth]{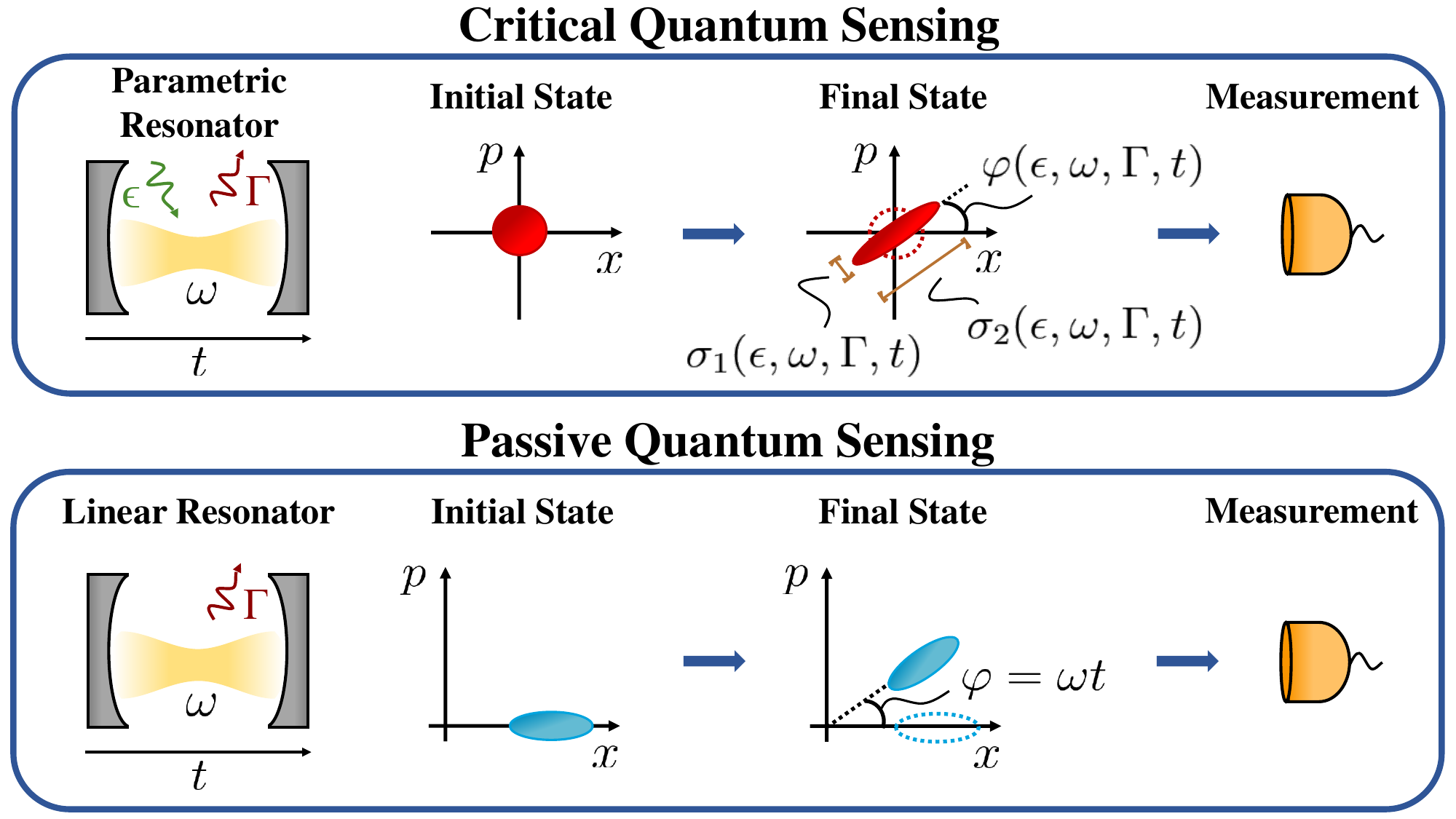}
\centering
\caption{{\bf Sketch of CQS and PQS strategies.} \textit{Top}: In CQS, the system, initially at the equilibrium with the environment, evolves according to \eqref{Kerr}. The final state is a squeezed thermal state with covariance matrix depending non-trivially on the system parameters. An optimal measurement is homodyne with an optimized angle, regardless of the system parameters. \textit{Bottom}: The initial state of PQS, an optimized displaced squeezed thermal state, acquires a phase shift \(\varphi=\omega t\) in the free time evolution. Here, to saturate the QFI, a non-linear measurement is needed in some parameter regimes. In both strategies, we consider interaction with a thermal environment as in \eqref{noise}.}
\label{sketch}
\end{figure}

Quantum advantage in sensing is defined in terms of the scaling of the achievable precision with respect to fundamental resources, such as system size and protocol duration time. Despite the critical slowing down, it has been shown~\cite{Rams2018} that CQS protocols implemented on many-body spin systems can achieve Heisenberg scaling~\cite{giovannetti2011advances} in both time and system size. This result has been recently extended~\cite{Garbe2020} to the class of finite-component PTs, which can take place in quantum resonators with atomic~\cite{Ashhab2013, hwang_quantum_2015,Puebla2017,Peng2019,Zhu2020} or Kerr~\cite{Bartolo2016,Felicetti2020,Minganti2023a,Minganti2023b} nonlinearities. In contrast to many-body spin systems, where criticality emerges in the limit of an infinite number of atoms, in finite-component models this thermodynamic limit is replaced with a rescaling of the physical parameters. While many-body spin systems become critical in the thermodynamic limit (infinite number of atoms), finite-component PTs are formally defined by a parameter-rescaling limit~\cite{hwang_quantum_2015,Felicetti2020} applied to a nonlinear bosonic system (infinite number of photons). 

On the one hand, finite-component PTs make it possible to implement CQS protocols with small-scale devices, such as parametric resonators~\cite{heugel2020_quantum,DiCandia2023,Rinaldi2021,petrovnin2023}, single trapped-ions~\cite{Ilias2023}, optomechanical~\cite{Bin2019,Tang2023} or magnomechanical~\cite{Wan2023} devices, spin impurities~\cite{mih23multiparameter} and Rabi-like systems~\cite{Ying2022,Xie2022,Lu2022}. On the other hand, finite-component PTs, as well as fully-connected systems~\cite{Garbe2022,Lambert04,Ribeiro07}, can be effectively described with minimal models, and so they provide a compelling theoretical framework to analyze CQS protocols with analytical or semi-analytical methods~\cite{Garbe2020,Salado2021,Chu2021,Garbe2022,Gietka2022,Gietka2022a,Garbe2022a,DiCandia2023,hotter2023}.
Recent theoretical efforts have been dedicated to the identification and design of optimal CQS protocols. It has been shown that the dynamical approach has a constant-factor advantage over static protocols~\cite{Chu2021,Garbe2022}. An apparent super-Heisenberg scaling can be achieved when focusing on a specific resource such as system size~\cite{Gietka2022,Gietka2022a} or time~\cite{Garbe2022a}. CQS protocols achieve quantum advantage also for global sensing using adaptive strategies~\cite{Montenegro2021,Salvia2023} in the driven-dissipative case with continuous measurements~\cite{Ilias2022,Yang2022} and in the multi-parameter case~\cite{Ivanov2020,DiFresco2022,mih23multiparameter}. Beyond the analysis of specific applicable protocols, in recent years, fundamental bounds on the quantum Fisher information (QFI)~\cite{Paris2009} have been derived~\cite{demkowicz2014using,demkowicz2017adaptive,zhou2021asymptotic,kurdzialek2022using,wan2022bounds}. Not only do they allow quick identification of which systems can benefit from quantum metrology, but they also clarify what should be considered a resource in metrology.

In this Letter, we fill several knowledge gaps in the understanding of criticality-enhanced protocols, by putting them in a general quantum metrology framework. We compare the performances of CQS and the standard quantum metrology approach, i.e., passive quantum sensing (PQS), in the frequency estimation task. We first consider only the system size as a resource. In the noiseless case, we find that, despite both strategies achieving Heisenberg scaling, optimal PQS outperforms CQS protocols by a constant factor. However, in the more realistic case of parameter estimation in dissipative dynamics, only CQS shows a \emph{quadratic} scaling of the single-shot QFI in the number of photons. This critical enhancement appears with the emergence of a transient regime from the unitary to the steady state dynamics, where the QFI grows. Such a regime can be arbitrarily long, and is not present in the absence of dissipation. 
Then, we consider both time and system size as resources, and we frame our results within the context of ultimate precision bounds. Here, there is a critical enhancement if preparation and/or measurement times are non-negligible. Finally, we show that our results stand also in the presence of thermal noise. 

Along the paper, we heavily use Gaussian quantum information methods for the solution of dynamics and for the computation of quantum and classical Fisher informations~\cite{Milburn,Serafini,vallone2019means}. To provide meaningful discussion, we may use approximations in the relevant regimes. However, all calculations are analytical, and their details are in the Supplemental Material (SM). See SM~\ref{sec:gaussian} for a summary of tools used. 

{\it Critical quantum sensing.---} We consider an idealized setting 
where the phenomenology of interest for CQS is described by the squeezing Hamiltonian,
\begin{align}\label{Kerr}
    H = \omega a^\dag a +\frac{\epsilon}{2}(a^2+a^{\dag 2})\,,
\end{align}
where $\epsilon$ is the squeezing parameter and $\omega=\omega_0+\delta \omega$ is the sum of a known frequency $\omega_0$ and an unknown, small, frequency shift $\delta \omega$ to be estimated. 
This minimal model can effectively describe~\cite{Garbe2022} the low-energy physics of a broad variety of criticalities emerging in: (i) finite-component systems such as the quantum Rabi model~\cite{hwang_quantum_2015,cai2021observation}, driven Kerr resonators~\cite{Bartolo2016,beaulieu2023observation,chen2023quantum}, ultrastrongly-coupled resonators~\cite{Felicetti2020}; and (ii) fully-connected models, such as the Dicke~\cite{Lambert04} and the Lipkin-Meshkov-Glick~\cite{Ribeiro07}. This system can be thought of as a Kerr resonator in the Gaussian approximation, i.e., in the limit of small Kerr non-linearity. In this limit, the system undergoes a second-order phase transition at the critical value 
$\epsilon=\epsilon_c=\sqrt{\omega^2+\Gamma^2}$~\cite{DiCandia2023}.
The effect of higher-order nonlinearities 
can be neglected until the photon number is sufficiently small, see SM~\ref{sec:critical}. The limits of validity of the approximation will be specific to each platform, and are not within the scope of this work.
We assume that the parameters $\omega_0$ and $\epsilon$ can be independently tuned, while $\delta \omega$ depends on some external field to be probed. To provide a practical example, the most direct implementation consists of a superconducting quantum resonator~\cite{beaulieu2023observation,chen2023quantum,zhong2013squeezing,KIPA1}, where $\epsilon$ corresponds to the intensity of an external parametric drive, $\omega_0$ is the detuning of the bare resonance frequency with respect to half the pump frequency, while $\delta\omega$ is directly proportional to an external magnetic flux.
We consider a coupling to a thermal bath, described by the Lindbladian 
\begin{align}\label{noise}
\mathcal{L}[\cdot] = &\Gamma (1+n_{B}) \left(2a\cdot a^\dag -\{a^\dag a,\cdot\}\right) \nonumber\\
\quad &+ \Gamma n_{B} \left(2a^\dag \cdot a -\{a a^\dag,\cdot\}\right)\,,
\end{align}
where $\Gamma\geq0$ is the environment-system coupling strength and $n_{B}$ is the effective temperature of the bath. 

We analyze a CQS protocol consisting of estimating the parameter $\delta \omega$ by choosing properly optimized values of $\omega_0$ and $\epsilon$, see Fig.~\ref{sketch}.  
Without loss of generality, we consider a constraint on the maximum average number of photons in the resonator, call it $N_{\rm max}$, that can theoretically be set arbitrarily large. This constraint is physically motivated as the model \eqref{Kerr} is the result of different approximations working for finite $N_{\rm max}$, such as the dispersive approximation when the resonator is coupled to an off-resonance qubit, or the Gaussian approximation~\cite{Garbe2022, DiCandia2023}.  

{\it Passive quantum sensing.---} PQS for the frequency estimation problem consists of initializing a linear resonator to a quantum state $\rho$, and letting it evolve according to the free Hamiltonian $H_0=\omega a^\dag a$ under the influence of noise in \eqref{noise}, see Fig.~\ref{sketch}. As in CQS, we assume that $\omega=\omega_0+\delta \omega$, where $\delta \omega$ is to be estimated. PQS assumes no active control over the resonator during the evolution, aside from choosing the interaction time. We consider the initial state generated with a generic unitary Gaussian operation applied to the state at the equilibrium with the environment, i.e.,  $\rho=D(\alpha)S(r)\rho_BS^\dag (r)D^\dag(\alpha)$, where $\rho_B$ is a thermal state with $n_{B}$ photons, $D(\alpha)$ and $S(r)$ are displacement and squeezing operations respectively, and the total number of photons is constrained to $N_{\rm max}=|\alpha|^2+(1+2n_{B})\sinh^2(r)+n_{B}$. 

{\it The noiseless case ($\Gamma=0$, $n_{B}=0$).---} Here, the QFI for estimating $\delta\omega$ with CQS is $I_{\rm cr}\sim [2N(t)+8N^2(t)/9]t^2$ for $\epsilon\to\epsilon_c$, where $N(t)\sim \omega_0^2t^2$. Details of the derivation can be found in SM~\ref{IVb}, where it is also shown that homodyne measurements saturates the QFI.
Notice that, with constraints on both $N_{\rm max}$ and the total time $T$, the optimal choice is to set $\omega_0=\sqrt{N_{\rm max}}/T$, so we can use all resources coherently. 
For PQS, by optimizing over Gaussian input states with $N_{\rm max}$ number of photons, we get $I_{\rm pas}= 8N_{\rm max}(1+N_{\rm max})t^2$. The optimal value is given by a squeezed-vacuum state, see SM~\ref{sec:passive}. Assuming $N(t)\leq N_{\rm max}$, $I_{\rm pas}$ is always larger than $I_{\rm cr}$ by a constant factor. This comes with no surprise, as PQS protocol is initialized with $N_{\rm max}$ photons while CQS with the vacuum. Here, the main message is that both protocols show quantum advantage, achieving the Heisenberg scaling $\propto (N_{\rm max} T)^2$. Notice that this analysis holds also for $\Gamma>0$, as long as $t\ll(N_{\rm max}\Gamma)^{-1}$. 

\begin{figure}[t!]
  \includegraphics[width=0.48
\textwidth]{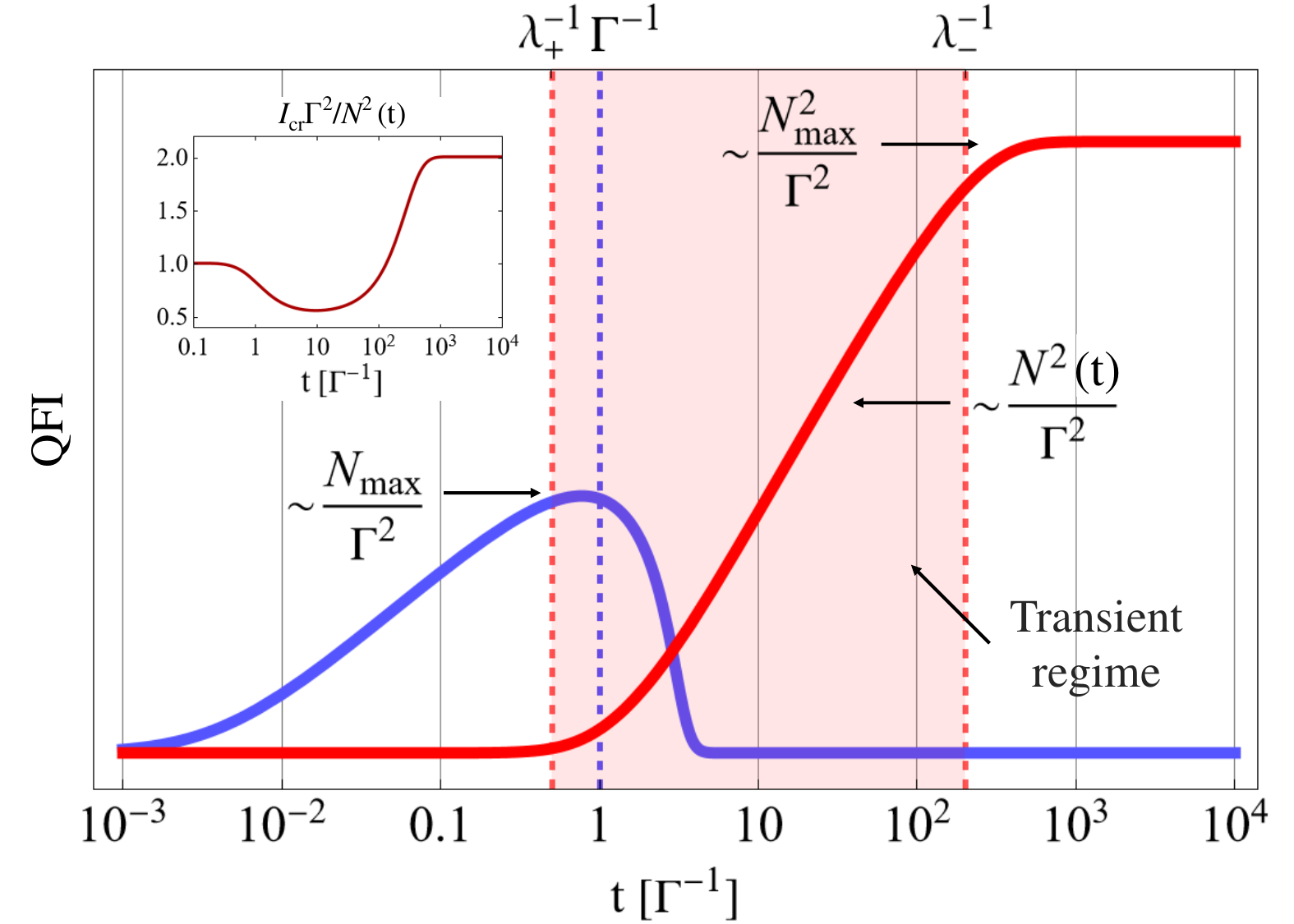}
\centering
\caption{{\bf Single-shot QFI.} Comparison of the single-shot QFI between PQS (blue) and CQS (red), at zero temperature, for \(\omega_0=\Gamma\) and \(N_{\rm max}=100\). 
The vertical axis has been rescaled as $\log(1+{\rm QFI})$ for better visibility.
For PQS, the optimal measurement time is \(t\simeq 0.8/\Gamma\). For CQS, the optimal $\epsilon$ is $\epsilon_{\rm opt}=\sqrt{\frac{2N_{\rm max}}{1+2N_{\rm max}}}\epsilon_c$. The critical enhancement is due to the emergence of the transient regime in the dissipative case, where the QFI grows with quadratic scaling with $N$ until reaching the steady state, see inset.}
\label{Fig2}
\end{figure}

What part of this quantum advantage will survive for longer times, where the effects of noise become significant? In the following, we first discuss the scaling of the {\it single-shot} QFI  with $N_{\rm max}$, therefore momentarily neglecting time as a resource. This will turn out to be useful for understanding the scaling of QFI with {\it both} $T$ and $N_{\rm max}$, which will be then related to ultimate precision bounds.

{\it Zero-temperature dissipative case ($\Gamma>0$, $n_{B}=0$).---} In the dissipative scenario, we recognize two different time scales for the critical dynamics, defined by the real parts of the Liouvillian eigenvalues $\lambda_{\pm}=\Gamma\pm\sqrt{\epsilon^2-\omega^2}$. Here, $\rm{Re}(\lambda_+)^{-1}$ is the time scale when the dynamic stops being effectively unitary, while $\rm{Re}(\lambda_-)^{-1}$ is the time scale to reach the steady state. For $\epsilon\leq\omega$ these times are equal, while for $\epsilon>\omega$ both $\lambda_\pm$ are real and different. This results in the emergence of a transient regime, see Fig.~\ref{Fig2}. Approaching the critical point $\epsilon\to\epsilon_c$ 
  makes the steady-state time diverge since $\lambda_-^{-1}\sim \Gamma/\epsilon_c(
\epsilon_c-\epsilon)$, so the transient regime can be arbitrarily long.

Let us switch to the problem of estimating $\delta\omega$. We consider $\epsilon>\omega_0$, and we work at $\omega_0=\Gamma$, which maximizes the QFI, see SM~\ref{IVb}. Also in the dissipative case, the optimal measurement is homodyne. From Fig.~\ref{Fig2}, we see that the interesting part is the transient regime, where the QFI is $I_{\rm cr}\gtrsim N^2(t)/2\Gamma^2$. The maximal QFI to $N^2$ rate is achieved at the steady state, where $I_{\rm cr}\simeq 2N^2(\infty)/\Gamma^2$, see inset of Fig.~\ref{Fig2}. The mean number of photons $N(t)$ increases monotonically in time and saturates at $N(\infty)=\epsilon^2/2(\epsilon_c^2-\epsilon^2)$. 
Looking for an optimal strategy with constraints on $N_{\max}$, since the optimal rate is at the steady state, the optimal choice of $\epsilon$ will be the one for which $N(\infty)=N_{\max}$, i.e., $\epsilon_{\rm opt}=\sqrt{\frac{2N_{\rm max}}{1+2N_{\rm max}}}\epsilon_c$. This analysis also shows that working close to the exceptional point $\epsilon\simeq\omega_0$ is a suboptimal choice, as at this point the number of photons is severely bounded.

For PQS, the QFI for estimating $\delta \omega$ is~\footnote{We set $\alpha$ and $r$ real, which is an optimal choice. This choice corresponds to squeezing the $p$ quadrature and displacing along the $x$ quadrature, as in Fig.~\ref{sketch}.}
\begin{align}\label{QFIpas0main}
I_{\rm pas} = &\,\left[\frac{4\alpha^2}{e^{-2r}+e^{2\Gamma t}-1}\right. \nonumber \\ 
\quad&\left.+\frac{e^{-2r}(e^{4r}-1)^2}{2e^{2r+4\Gamma t}+(e^{2r}-1)^2(e^{2\Gamma t}-1)}\right]t^2\,.
\end{align}
Let us consider $t\gtrsim (N_{\rm max}\Gamma)^{-1}$. Under the condition $e^{2r}\gg e^{4\Gamma t}/(e^{2\Gamma t}-1)$, we get the simple expression
\begin{align}\label{QFIpasopt}
I_{\rm pas} \simeq \frac{4N_{\rm max}t^2}{e^{2\Gamma t}-1}\,.
\end{align}
The condition on $r$ can be easily satisfied also at finite $N_{\rm max}$ if $\Gamma t$ is not too large.
One can see that, to optimize the QFI, the exact amount of squeezing is not crucial as long as it guarantees the condition on $r$. The QFI \eqref{QFIpasopt} is optimal for $t\simeq 0.8/\Gamma$, for which $I_{\rm pas}=O(N_{\max}/\Gamma^2)$, see Fig.~\ref{Fig2}. We should notice that the first term in \eqref{QFIpas0main} corresponds to the Fisher information for homodyne measurement of the $p$ quadrature, which saturates the QFI for $t\gtrsim (N_{\rm max}\Gamma)^{-1}$ already when $N_{\rm max}\gtrsim 10^3$, see SM~\ref{sec:passive}.

We see a difference in scaling in the number of photons between CQS and PQS as $I_{\rm cr}=O(N_{\rm max}^2/\Gamma^2)$, while $I_{\rm pas}=O(N_{\rm max}/\Gamma^2)$. This is the signature of the critical enhancement. It is also clear that this enhancement emerges from the splitting of the real part of the Liouvillian eigenvalues for $\epsilon>\omega_0$, which allows $\lambda_-^{-1}$ to be arbitrarily large for $\epsilon$ approaching $\epsilon_c$, as $\lambda_-^{-1}=O(N_{\rm max}/\Gamma)$. As we will see in the next section, when considering {\it both} time and system size as a resource, the optimal scaling for the QFI is $O(N_{\rm max}t/\Gamma)$, see Eq.~\eqref{eq:fundbound}. CQS allows for the coherent use of time $t\sim \lambda_-^{-1}$, from which the quadratic scaling for the QFI follows. In the absence of the transient regime, i.e., for $\epsilon<\omega_0$, there is a single time scale for the dynamics given by $t\sim\Gamma^{-1}$, so the QFI scales linearly. This bound holds for any initial state, and, therefore, explains also why PQS shows a linear scaling for the QFI (see Fig.~\ref{Fig2}).

\begin{figure}[t!]
  \includegraphics[width=0.47
\textwidth]{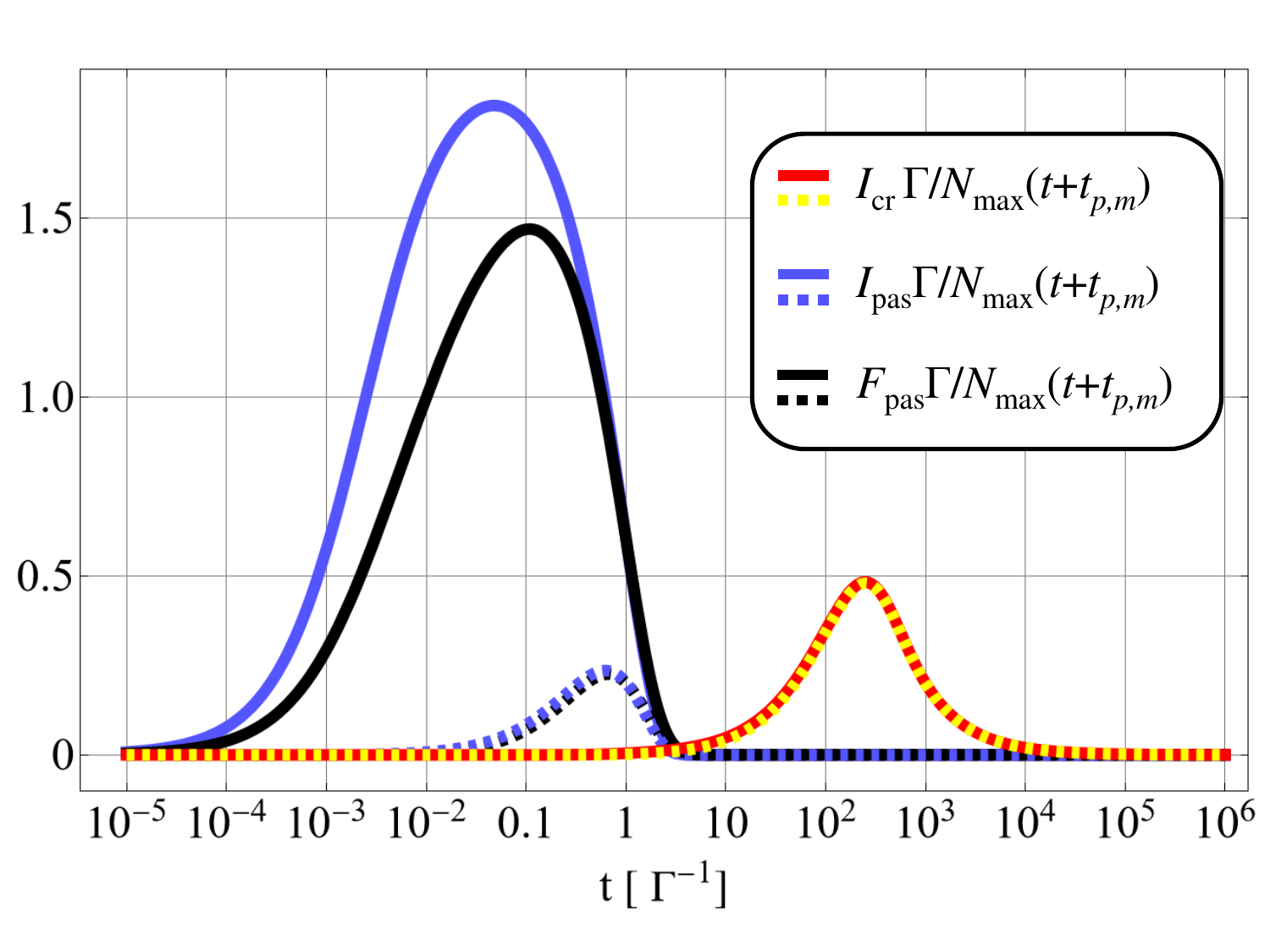}
\centering
\caption{{\bf QFI rate.} Comparison of the ratios \(I_{\rm pas,cr}/N_{\rm max}(t+t_{p,m})\) for the PQS (blue) and CQS (red), at zero temperature, for $t_{p,m}=0$ (solid lines) and $t_{p,m}=2/\Gamma$ (dashed lines). In black, we draw the same type of plot for the Fisher Information of homodyne measurement in PQS. Here, we set $N_{\rm max}=100$,\, \(\epsilon=\epsilon_{\rm opt},\,\omega_0=\Gamma\). By neglecting preparation and measurement time, the passive strategy is fundamentally optimal for large enough $N_{\rm max}$, as it saturates the ultimate precision bounds. Even for finite $N_{\rm max}$, it performs significantly better than the critical strategy. However, by considering $t_{p, m}>0$, while the QFI rate is significantly reduced for PQS, it remains essentially unchanged for CQS. In this framework, there is a critical enhancement. 
}
\label{Fig3}
\end{figure}

{\it Relation to ultimate precision bounds.---} So far the analysis has been carried out considering $N_{\rm max}$ alone as a resource. In the occurrence of losses, it is not possible to use the entire time resource coherently. However, as QFI arises linearly with the number of repetitions and a shorter time of single realization allows for a bigger number of repetitions, for fair comparison, we should still treat both $N_{\rm max}$ and total time $T$ as a resource. Then, to use them optimally, one should divide the total time $T$ into smaller parts $t_{\rm opt}={\rm argmax}_{t} I(t)/t$, and in total time $T$ perform $M=T/t_{\rm opt}$ repetitions.
For passive strategies, this leads to $MI_{\rm pas}\sim 2N_{\rm max}T/\Gamma$ for $N_{\rm max}\gg1$, where $t_{\rm opt}$ decreases with increasing $N_{\rm max}$, see SM~\ref{optimal} for details.

To analyze the critical protocol in this framework, note that, as in the transient regime the number of photons increases linearly with time, for $\omega_0=\Gamma$ and close to the criticality the time of a single repetition scales roughly as $\lambda_-^{-1}\simeq 2N_{\max}/\Gamma$. Therefore, the number of repetitions decreases with $N_{\max}$ as $M\simeq T\Gamma/2N_{\max}$, so the
scaling $I_{\rm cr}=O(N_{\max}^2/\Gamma^2)$ translates to $MI_{\rm cr}=O(N_{\max}T/\Gamma)$, as in passive strategy. It is also worth emphasizing that, to obtain the scaling $\propto N_{\max}T/\Gamma$ of the QFI, no quantum resources are needed, i.e., a protocol based on a coherent state with single repetition time $1/\Gamma$ and homodyne detection achieves this scaling as well.

Can this scaling be improved in any way? By applying results from~\cite{demkowicz2017adaptive,wan2022bounds}, we show that the QFI for the estimation of the frequency of the 
cavity coupled to the thermal bath is fundamentally bounded by (see SM~\ref{sm:fundamental}):
\begin{equation}\scalebox{0.96}{
\label{eq:fundbound}\(\displaystyle
I^{\rm total}_{\rm cr,pas}\leq \int_0^T \frac{2N(t)}{\Gamma(1+2n_B-\frac{n_B}{N(t)+1})}dt\lesssim \frac{2N_{\max}T}{\Gamma(1+2n_B)}\,,\)}
\end{equation}\\
where the second inequality holds for $N_{\max}/n_B\gg 1$. Here with the superscript ''total" we stress the fact that the bound already includes the possibility of dividing the total time $T$ into smaller parts and perform measurements between them (QFI scales linearly with the number of repetitions). While in this paper we discuss in detail the protocol based on phase transition in the occurrence of squeezing Hamiltonian, the above bound remains valid for any other metrological strategy, including all kinds of criticality, adaptiveness, partial measurements etc. Note that, while optimal PQS saturates the bound in the limit of large $N_{\max}$, the CQS cannot perform as well, since the number of photons arises from $0$. Therefore, after averaging, it needs to be strictly smaller than $N_{\max}$. Where, then, does the advantage of CQS manifest itself?

The ultimate bound \eqref{eq:fundbound} is derived by neglecting preparation and measurement time $t_{p,m}$. In many experiments, this is an unrealistic assumption. For instance, to initialize a {\rm linear} resonator to a squeezed state with $N_{\rm max}$ photons, one way involves pumping the resonator with a squeezed signal. Assuming the favorable situation that $\Gamma$ represents the coupling with the preparation line, it then takes $O(\log(N_{\rm max})/\Gamma)$ time to prepare the cavity, see SM~\ref{timechargedischarge}. Discharging the resonator, essential to perform measurements on the output modes, requires the same time. Generally speaking, a more meaningful way to approach the problem is to divide $T$ in $t_{\rm opt}={\rm argmax}_{t}I(t)/(t+t_{p,m})$ parts. In Fig.~\ref{Fig3}, we show that, already for $t_{p,m}\simeq 2/\Gamma$, i.e., a time $1/\Gamma$ each for measuring and preparing the state, PQS performance is largely reduced while CQS performance remains virtually untouched. This is because in CQS the single-shot QFI achieves its maximum at a time much larger than $\Gamma^{-1}$, so $t_{p,m}$ is negligible.
This leaves space for independent exploration by considering specific implementations of the protocols. For instance, preparation and measurement of the field {\it outside} the resonator can be further analyzed using time-dependent input-output theory.

{\it Finite-temperature dissipative case ($\Gamma>0$, $n_{B}>0$).---} A similar analysis can be performed for arbitrary temperature. For the dynamics, we consider the critical system starting from a thermal state $\rho_B$ with $n_{B}$ photons and consider $\epsilon>\omega_0$. For the same values of $\epsilon,\,\omega_0$, the Liouvillian eigenvalues are unchanged, so the unitary and steady-state time scales are the same. Moreover, also \(\epsilon_c\) and $I_{\rm cr}(t)$ are left unchanged. However, the mean number of photons at any time is $(1+2n_B)$ times bigger than in the zero-temperature case. It means that the same value of QFI would be obtained if the constraint for the number of photons would be also rescaled to $N_{\max}'=(1+2n_B)N_{\max}$. The same holds also for the passive strategy, see SM~\ref{noiseQFI}. Both protocols are therefore robust to thermal noise in the same way, in accordance to the bound \eqref{eq:fundbound}.

{\it Beyond the critical point.---} Lastly, we shall discuss whether it is possible to get an enhancement by exploiting the dynamics of a fast quench of the system, i.e., working at $\epsilon> \epsilon_c$, as proposed in Ref.~\cite{Gietka2022}. For $\epsilon> \epsilon_c$, the number of photons grows exponentially in time, as $N(t)\sim e^{2\sqrt{\epsilon^2-\epsilon_c^2} t}/4$. Since the QFI is polynomial in the number of photons, also the QFI increases exponentially in time. One may then conclude that this strategy offers a great advantage. However, an analysis based on imposing a constraint on the number of photons in the resonator reveals that this strategy is suboptimal.

Consider for instance the noiseless case, see SM~\ref{IVb}. Here, $I_{\rm cr}^{\epsilon>\epsilon_c}(t)\sim 4N^2(t)/(\epsilon^2-\epsilon_c^2)$ for $t \sqrt{\epsilon^2-\epsilon_c^2}\gg1$. The optimal choice of $\epsilon$, allowing for coherent use of all resources, under the photon number constraint $N(T)=N_{\max}$, is $\epsilon^2\simeq\epsilon_c^2+\log^2(4N_{\max})/4T^2$, which leads to $I_{\rm cr}^{\epsilon>\epsilon_c}=O(N_{\max}^2T^2/\log^2N_{\max})$. So, contrary to the case below the critical point, Heisenberg scaling with all resources is not possible at all. 

{\it Conclusions.---}  
We have compared passive quantum sensing strategies with protocols exploiting the dynamics of driven-dissipative critical systems. We have identified relevant frameworks in which critical quantum sensing outperforms passive quantum sensing for parameter estimation task, in open quantum systems at arbitrary temperature. The considered minimal model describes the critical behavior of a broad class of systems, including finite-component phase transitions~\cite{hwang_quantum_2015,Felicetti2020,cai2021observation,beaulieu2023observation,chen2023quantum} and fully-connected models~\cite{Garbe2022,Lambert04,Ribeiro07}. PTs of this kind have been already observed with controllable atomic~\cite{baumann2010dicke,cai2021observation} and solid-state~\cite{beaulieu2023observation,chen2023quantum,zhong2013squeezing,KIPA1} quantum technologies. For critical models that do not belong to the considered class, such as many-body spin models, our paper still provides a method to make a comparison with the ultimate precision performance.
As the critical enhancement appears for dissipative systems and is robust against thermal noise and measurement/preparation time, our analysis paves the way for the development of {\it practical} critical quantum sensors in these experimental settings. Indeed, in some experimental contexts CQS sensing protocols can be even simpler to implement than standard sensing strategies, as the initialization does not depend on the prior and the optimal measurement is a simple homodyne detection in all regimes.

{\it Acknowledgements.---} We thank Rafa{\l} Demkowicz-Dobrza{\'n}ski and Pavel Sekatski for useful comments on fundamental bound on QFI. We acknowledge financial support from the Academy of Finland, grants no. 353832 and 349199, from the U.S. DoE,
National Quantum Information Science Research Centers, Superconducting
Quantum Materials and Systems Center (SQMS) under contract number
DE-AC02-07CH11359, from EU H2020 Quant ERA ERA-NET Cofund in Quantum
Technologies QuICHE under Grant Agreement 731473 and 101017733, from
the PNRR MUR Project PE0000023-NQSTI, from the National Research
Centre for HPC, Big Data and Quantum Computing, PNRR MUR Project
CN0000013-ICSC, and from PRIN2022 CUP 2022RATBS4.

\bibliography{biblio2.0.bib}

\onecolumngrid
\newpage

\begin{center}
{\huge \bf{Supplemental Material to:\\ ``Optimality and Noise-Resilience of Critical Quantum Sensing''}}
\end{center}
\renewcommand{\thesection}{\Roman{section}}
\numberwithin{equation}{section}
This Supplemental Material shows details about the claims in the paper. We have used Mathematica to perform all calculations. Most of the formulas are too large to be written in the text. In such a case, we provide insight in the form of asymptotic expansion in relevant regimes. 

\section{Gaussian Formalism}
\label{sec:gaussian}
Gaussian states are fully characterized by the first-moment vector ${\bf v}$ and the covariance matrix ${\bf \Sigma}$. For a Gaussian mode $a$, with $[a,a^\dag ]=1$, these objects are defined as
\begin{align}
   {\bf v}&=\begin{pmatrix}
        \langle x\rangle\\
       \langle p\rangle
    \end{pmatrix}\,, \label{fmv1}\\
     {\bf \Sigma}&= \begin{pmatrix}
        2\langle x^2\rangle-2\langle x\rangle^2& \langle\{x,p\}\rangle-2\langle x\rangle\langle p\rangle\\
        \langle\{x,p\}\rangle-2\langle x\rangle\langle p\rangle&2\langle p^2\rangle-2\langle p\rangle^2
    \end{pmatrix} \label{covmatr1} \,,
    \end{align}
where $x= \left(a+a^\dag\right)/\sqrt{2}$ and $p=-i \left(a-a^\dag\right)/\sqrt{2}$.

Given a manifold of Gaussian states ${\rho_\theta}$ dependent on a parameter $\theta$, the Quantum Fisher Information (QFI) computed in $\theta_0$ quantifies the maximal precision achievable to estimate $\theta$ when its value is close to $\theta_0$. This can be seen via the quantum Cramer-Rao bound, which states that $\Delta \theta^2|_{\theta\simeq\theta_0}\geq 1/MI(\theta_0)$. Here, $M$ is the number of repetitions and $I(\theta_0)= \lim_{d\theta\to 0}\frac{8}{d\theta^2}\left[1-D(\rho_{\theta_0},\rho_{\theta_0-d\theta})\right]$, with $D[\rho,
\sigma]=\left[{\rm Tr}\,\left(\sqrt{
\rho\sqrt{
\sigma}\rho}\right)\right]^2$ the fidelity, is the
QFI~\cite{Paris2009}. For single-mode Gaussian states, this can be computed as 
\begin{equation}\label{QFIGaussian}
    I(\theta_0)=\frac{1}{2}\frac{\Tr\left[\left({\bf \Sigma}^{-1}\partial_{\theta}{\bf \Sigma}\right)^2\right]}{1+\mu^2}+\frac{2\left(\partial_{\theta}\mu\right)^2}{1-\mu^4}+2\left(\partial_{\theta}{\bf v}\right)^T{\bf \Sigma}^{-1}\partial_{\theta} {\bf v}\,,
\end{equation}
where \(\mu=1/\sqrt{\Det[{\bf \Sigma}]}\) is the purity of the system state and the derivatives are computed in $\theta_0$~\cite{Serafini}.

When fixing the POVM $\{\Pi(x)\}$, the maximal achievable precision is quantified by the classical Cramer-Rao bound, i.e., $\Delta \theta^2|_{\theta=\theta_0}\geq 1/MF(\theta_0)$. Here, $F(\theta_0)=\int_{-\infty}^{\infty}\left(\partial_\theta \log (p_\theta(x))|_{\theta=\theta_0}\right)^2p_{\theta_0}(x)dx$, with $p_\theta(x)={\Tr}\left(\rho_\theta \Pi(x)\right)$, is the Fisher Information (FI).  If we measure a quadrature $x(\psi)=(e^{-i\psi}a +e^{i\psi} a^\dag)/\sqrt{2}$ of a state belonging to the manifold of Gaussian states ${\rho_\theta}$ , with \(\psi\in[0,2\pi)\), the FI can be readily written as 
\begin{align}\label{FI}
F(\theta_0)=\frac{4S(\psi)\left[\partial_{\theta}\langle x(\psi)\rangle\right]^2+\left[\partial_{\theta} S(\psi)\right]^2}{2S^2(\psi)}\,,
\end{align}
where \(S(\psi)=\cos^2{(\psi)}\Sigma_{11}+\sin^2{(\psi)}\Sigma_{22}-\sin{(2\psi)}\Sigma_{12}\), $\Sigma_{ij}$ are the elements of the covariance matrix ${\bf \Sigma}$, and the derivatives are computed in $\theta_0$.

Alternatively, a parameter may be estimated from the mean number of photons \(N\). Assuming ${\bf v}=0$, we have $N=\frac{1}{4}(\Sigma_{11}+\Sigma_{22})-\frac{1}{2}$ and $\Delta^2 N=\frac{1}{16}(\Sigma_{11}^2+\Sigma_{22}^2+2\Sigma_{12}^2)-\frac{1}{4}$~\cite[Sec. IV, A]{vallone2019means} [note the difference by a factor $2$ in the definition of the covariance matrix], which leads to signal-to-noise ratio $S(\theta)=\frac{|\partial_\theta\langle N\rangle_{\theta}|^2 }{\Delta^2 N_\theta}$.

\section{The Critical System}
\label{sec:critical}
\noindent In this section, we discuss details on the critical model, including the calculation of the time-dependent solution of the mode, explicit evaluation of the first-moment vector and the covariance matrix, computation of the number of photons and recognition of different time scales of the dynamics.
\subsection{The Model}
\label{IIA}
We consider a Kerr resonator Hamiltonian in the Schr\"odinger's picture defined as
\begin{align}\label{HSchro}
H_S = (\omega_r+\delta \omega) a^\dag  a +\frac{\epsilon}{2}\left(e^{-i\omega_p t}a^2+e^{i\omega_pt}a^{\dag 2} \right) +\chi a^{\dag 2}a^2\,,
\end{align}
where $\omega_r+\delta \omega$ is the resonator frequency, $\omega_p$ is the pump frequency and $\chi>0$ is the Kerr non-linearity. The model is valid for $|\omega_r-\omega_p/2|\ll \omega_r$ (assuming $\delta \omega\ll \omega_r$). We work in the frame rotating with $\omega_p/2$. The Hamiltonian in this picture is 
\begin{align}\label{CQHam}
H = \omega a^\dag a +\frac{\epsilon}{2}\left(a^2+a^{\dag 2}\right) +\chi a^{\dag 2}a^2\,,
\end{align}
where $\omega=\omega_0+\delta \omega$ and  $\omega_0=\omega_r-\omega_p/2$. In the context of quantum parameter estimation, we will assume $\epsilon$, $\omega_r$ and $\omega_p$ (therefore also $\omega_0$) to be known and $\delta \omega$ to be estimated. Notice that the parameter $\omega_0$ can be tuned by changing the pump frequency $\omega_p$, as long as $|\omega_r-\omega_p/2|\ll \omega_r$ is respected. In the main text, we work with $\omega_0=\Gamma$, which means that $\Gamma\ll\omega_r$ is assumed.
We consider the Lindbladian modeling a thermal environment 
\begin{align}\label{noise_suppl}
\mathcal{L}[\cdot] = \Gamma (1+n_{B}) \left(2a\cdot a^\dag -\{a^\dag a,\cdot\}\right)+ \Gamma n_{B} \left(2a^\dag \cdot a -\{a a^\dag,\cdot\}\right)\,,
\end{align}
where $\Gamma\geq0$ is coupling with the bath and $n_{B}\geq0$ is the effective temperature.

We observe that, for $\chi=0$, the Hamiltonian is bounded from below up to a certain value of $\epsilon$, i.e., \(\epsilon_c=\sqrt{\omega^2+\Gamma^2}\). Mathematically, the Kerr non-linearity term $\chi a^{\dag 2}a^2$ regularizes the model in a way that makes it physical for all system parameter values. For $\chi\to0$, the system undergoes a second-order phase transition~\cite{DiCandia2023} for a critical value of $\epsilon$. In this paper, we work in the Gaussian approximation and set $\chi=0$. This approximation holds until the number of photons is bounded by roughly $N\lesssim O(\epsilon/\chi)$. Notice that a very weak Kerr non-linearity is achievable in, e.g., Kinetic Inductance Parametric Amplifiers, where the ratio $\epsilon/\chi$ can be as large as $10^{8}$~\cite{KIPA1}.

\subsection{Time-dependent solution}

\subsubsection{Solution of the Langevin equation}
\begin{figure}[t!]
  \includegraphics[width=1
\textwidth]{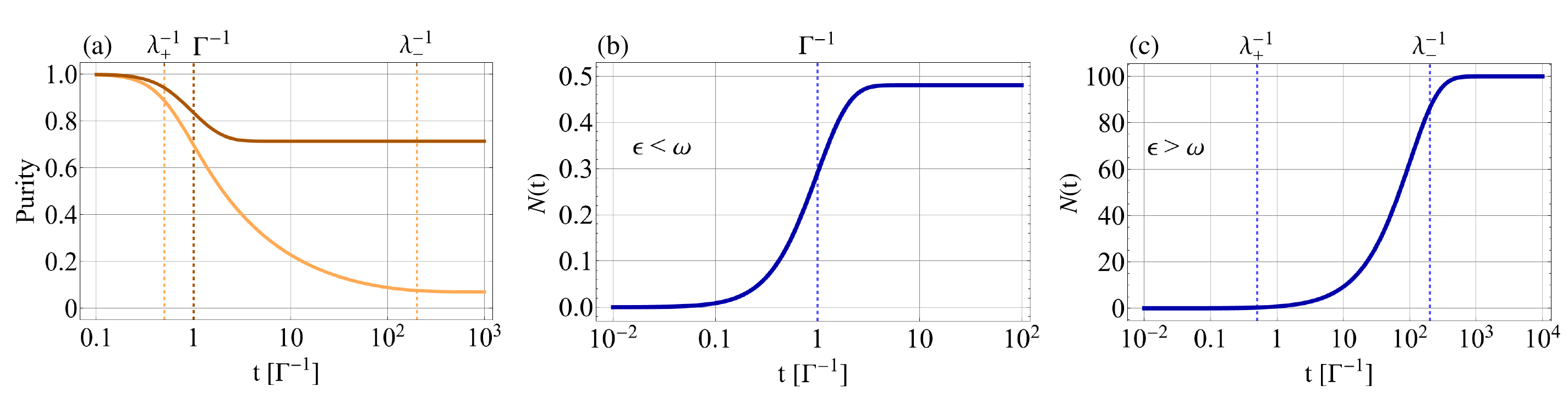}
\centering
\caption{\textbf{Purity and mean photon number.} {\bf (a)} We plot the purity of the CQS system state, for \(\epsilon>\omega\) (orange) and \(\epsilon<\omega\) (brown). Here, it is clear that ${\rm Re}(\lambda_+)^{-1}$ is the time scale for the purity to drop. Notice that, for \(\epsilon<\omega\), ${\rm Re}(\lambda_+)^{-1}={\rm Re}(\lambda_-)^{-1}=\Gamma^{-1}$.  {\bf (b,c)} Mean photon number for $\epsilon<\omega$ and $\epsilon>\omega$. For $\epsilon<\omega$ the time scale to reach the steady state coincides with the time scale for the purity to drop. For \(\epsilon>\omega\), there are two distinct timescales, and the time scale to reach the steady state is \(\lambda^{-1}_-\). To obtain the plot, we set \(\omega=\Gamma\), \(n_{B}=0\), \(\epsilon=0.99\Gamma\) (when $\epsilon<\omega$), and \(\epsilon=0.9975\epsilon_c\) (when $\epsilon>\omega$).}
\label{Fig4}
\end{figure}

We study the dynamics of the cavity mode \(a(t)\) by solving the associated Langevin equation~\cite{Milburn}
\begin{equation}\label{diffeqtwosided}
    \Dot{\textbf{a}}(t)=A\textbf{a}(t)-\sqrt{2\Gamma}\textbf{b}(t)\,,
\end{equation}
where we have introduced
\begin{equation}
    \textbf{a}(t)=\begin{pmatrix}
        a(t)\\a^\dag(t)
    \end{pmatrix}\,,\,\,\,\textbf{b}(t)=\begin{pmatrix}
        b(t)\\b^\dag(t)
    \end{pmatrix}
    \,,\,\,\,A=\begin{pmatrix}
        -i\omega-\Gamma&-i\epsilon\\i\epsilon&i\omega-\Gamma
    \end{pmatrix}\,,
\end{equation}
and \(b(t)\) is a thermal mode satisfying \(\langle b(t')b^\dag(t)\rangle=(1+n_{B})\delta(t'-t)\). Eq.~\eqref{diffeqtwosided} can be solved with direct integration, obtaining
\begin{align}\label{cavitytwosided}
a(t)&=c_1(t)a(0)+c_2(t)a^\dagger(0)-\sqrt{2\Gamma}\int_{0}^{t}c_1(t-\tau)b(\tau)d\tau-\sqrt{2\Gamma}\int_{0}^{t}c_2(t-\tau)b^\dagger(\tau)d\tau\,,
\end{align}
where the coefficients \(c_1,\,c_2\) are defined as
\begin{align}
    c_1(x)&=\left(\frac{1}{2}-\frac{i\omega}{2\sqrt{\epsilon^2-\omega^2}}\right)e^{-\lambda_-x}+\left(\frac{1}{2}+\frac{i\omega}{2\sqrt{\epsilon^2-\omega^2}}\right)e^{-\lambda_+x}\,,\label{c1}\\
    c_2(x)&=\frac{-i\epsilon}{2\sqrt{\epsilon^2-\omega^2}}\left(e^{-\lambda_-x}-e^{-\lambda_+x}\right)\label{c2}\,,
\end{align}
and $\lambda_\pm$ are the eigenvalues of the matrix \((-A)\), i.e., \(\lambda_\pm=\Gamma\pm\sqrt{\epsilon^2-\omega^2}\). We will see that the real parts of $\lambda_-$ and $\lambda_+$ define two different time scales. Indeed, ${\rm Re}(\lambda_+)^{-1}$ is the characteristic time for the purity decay, and ${\rm Re}(\lambda_-)^{-1}$ is the characteristic time to reach the steady state, see Fig.~\ref{Fig4}.

\subsubsection{First-moment vector and covariance matrix}
We assume the cavity at $t=0$ to be at equilibrium with the environment, so $a(0)$ is a thermal state with $n_{B}$ photons. Therefore, the mode $a(t)$ is Gaussian at all times and is fully characterized by its first-moment vector ${\bf v}$ and covariance matrix ${\bf \Sigma}$. 
The quantities ${\bf v}$ and ${\bf \Sigma}$, defined in Eqs.~\eqref{fmv1}-\eqref{covmatr1}, may be written as:
\begin{align}
   {\bf v}(t)&=\begin{pmatrix}
        (\langle a(t)\rangle +\langle a^\dag(t)\rangle)/\sqrt{2}\\
       -i(\langle a(t)\rangle-\langle a^\dag(t)\rangle)/\sqrt{2}
    \end{pmatrix}\,, \label{fmv}\\
     {\bf \Sigma}(t)&= \begin{pmatrix}
        2\langle a^\dag(t)a(t)\rangle+\langle a(t)a(t)\rangle+\langle a^\dag(t)a^\dag(t)\rangle+1& i(\langle a^\dag(t)a^\dag(t)\rangle-\langle a(t)a(t)\rangle)\\
        i(\langle a^\dag(t)a^\dag(t)\rangle-\langle a(t)a(t)\rangle) & 2\langle a^\dag(t) a(t)\rangle-\langle a(t)a(t)\rangle-\langle a^\dag(t)a^\dag(t)\rangle +1
    \end{pmatrix} \label{covmatr} \,,
    \end{align}
and then can be directly computed from \eqref{cavitytwosided} using the relations
\begin{align}
\langle a(t)\rangle &= 0\,,\\
\langle a^2(t) \rangle &= (2n_{B}+1)c_1(t)c_2(t)+(4n_{B}+2)\Gamma \int_0^tc_1(t-\tau)c_2(t-\tau)d\tau\,, \\
\langle a^\dag (t) a(t)\rangle &= |c_2(t)|^2+ 2\Gamma\int_0^t |c_2(t-\tau)|^2d\tau+n_{B}\left\{|c_1(t)|^2+|c_2(t)||^2+2\Gamma\int_0^t\left[|c_1(t-\tau)|^2+|c_2(t-\tau)|^2\right]d\tau\right\}\,,
\end{align}
with $c_1(t)$, $c_2(t)$ defined in \eqref{c1}-\eqref{c2}. Notice that the first-moment vector is zero at all times, therefore the state is characterized solely by the covariance matrix. As all integrals above are solvable, that gives an analytical formula for a covariance matrix, which is, however, too long to put it here.

\subsubsection{The steady state}
At the steady state, one gets the covariance matrix 
\begin{align}
\label{eq:covsteady}
{\bf \Sigma}_{ss}&= \frac{1+2n_{B}}{\epsilon_c^2-\epsilon^2} \begin{pmatrix}
   \epsilon^2_c-\omega\epsilon     & -\Gamma \epsilon \\
  -\Gamma \epsilon      &  \epsilon^2_c+\omega\epsilon 
    \end{pmatrix}\,,
\end{align}
where $\epsilon_c=\sqrt{\omega^2+\Gamma^2}$. In addition, the number of photons at the steady state is 
\begin{align}\label{Nss}
N(\infty)= \frac{\epsilon^2+2n_{B}\epsilon_c^2}{2(\epsilon_c^2-\epsilon^2)} = \frac{(1+2n_B)\epsilon^2}{2(\epsilon_c^2-\epsilon^2)}+n_B\,,
\end{align}
which diverges for $\epsilon\to \epsilon_c$. Here, we emphasize that for $\epsilon>\epsilon_c$, there is no steady state.

\subsection{Mean number of photons}\label{Ntcr}

\subsubsection{Zero-temperature case ($n_{B}=0$)}
In this case, we can derive short formulas for the average number of photons. We have that 
\begin{align}
|c_2(t)|^2&=\frac{\epsilon^2}{4|\epsilon^2-\omega^2|}\left[e^{-2{\rm Re}(\lambda_-)t}+e^{-2{\rm Re}(\lambda_+)t}-2e^{-2\Gamma t}\cos\left(2{\rm Im}\left(\sqrt{\epsilon^2-\omega^2}\right)\cdot t\right)\right]\,,\\
\int_0^t |c_2(t-\tau)|^2d\tau &= \frac{\epsilon^2}{4|\epsilon^2-\omega^2|}\left\{\frac{1-e^{-2{\rm Re}(\lambda_-) t}}{2{\rm Re}(\lambda_-)}+\frac{1-e^{-2{\rm Re}(\lambda_+)t}}{2{\rm Re}(\lambda_+)}+\right. \nonumber\\
\quad &\quad \left.-\frac{\Gamma-e^{-2\Gamma t}\Gamma\cos\left(2{\rm Im}\left(\sqrt{\epsilon^2-\omega^2}\right)\cdot t\right)+e^{-2\Gamma t}{\rm Im}\left(\sqrt{\epsilon^2-\omega^2}\right) \sin\left(2{\rm Im}\left(\sqrt{\epsilon^2-\omega^2}\right) \cdot t\right)}{[{\rm Im}\left(\sqrt{\epsilon^2-\omega^2}\right)]^2+\Gamma^2}\right\}\,.
\end{align}
Let us consider the two cases $\epsilon<\omega$ and $\epsilon>\omega$ separately.

For $\epsilon<\omega$, we have that ${\rm Re}(\lambda_-)={\rm Re}(\lambda_+)\equiv \Gamma$ and ${\rm Im}(\sqrt{\epsilon^2-\omega^2})=\sqrt{\omega^2-\epsilon^2}$. Therefore,
\begin{align}
N_{\epsilon<\omega}(t)&= \frac{\epsilon^2}{2(\epsilon^2_c-\epsilon^2)}\left\{1-e^{-2\Gamma t}\left[\cos\left(2\sqrt{\omega^2-\epsilon^2}\cdot t\right)+\frac{\Gamma }{\sqrt{\omega^2-\epsilon^2}}\sin\left(2\sqrt{\omega^2-\epsilon^2}\cdot t\right)\right]\right\}\,.
\end{align}

For $\epsilon>\omega$, we have that $\lambda_\pm$ are real and ${\rm Im}(\sqrt{\epsilon^2-\omega^2})=0$. Therefore,
\begin{align}
N_{\epsilon>\omega}(t)&= \frac{\epsilon^2
}{2(\epsilon_c^2-\epsilon^2)}\left[1-\left(\frac{\Gamma}{\sqrt{\epsilon^2-\omega^2}}+1\right)e^{-2\lambda_-t}+\left(\frac{\Gamma}{\sqrt{\epsilon^2-\omega^2}}-1\right)e^{-2\lambda_+t}\right]\,.
\end{align}

It is clear that, for $\epsilon<\omega$, there is only a single time-scale given by $\Gamma^{-1}$, while, for $\epsilon>\omega$, we have two different time scales, i.e., $\lambda_-^{-1}$ and $\lambda_+^{-1}$. Looking at the purity in Fig.~\ref{Fig4}, we can say that $\lambda_+^{-1}$ is the unitary time, i.e., the characteristic time for the purity to decay. Instead, $\lambda_-^{-1}$ is the characteristic time to reach the steady state.

\subsubsection{Finite-temperature case}\label{NumnT}

For arbitrary values of $n_{B}$, the expression for the number of photons is quite large. However, one can gain some insight by doing an asymptotic analysis. At the steady state, we have that $N_{n_B}(\infty) = (1+2n_{B})N_{n_{B}=0}(\infty)+n_B$, as shown in \eqref{Nss}. At finite time, we first expand $N_{n_{B}=0}(t)$ at the first order around $\epsilon= \epsilon_c$. This expansion holds for $t\ll \lambda_-^{-1}$. Then, by taking the series for large \(n_{B}\), we get
\begin{align}\label{NTemp}
N_{n_B}(t)\simeq (1+2n_{B})N_{n_{B}=0}(t)+n_{B}\,.
\end{align}
Numerical evaluation shows that \eqref{NTemp} is indeed an accurate approximation for any value of $n_{B}$.

\subsection{Noiseless case ($\Gamma=0$, $n_B=0$)}

Simple and compact formulas can be obtained for the noiseless case $\Gamma=0$. Note that then $\epsilon_c=\omega$. In such a case, the time-dependent covariance matrix is given by the formula:
\begin{equation}
\label{eq:covgamma0}
    {\bf \Sigma}(t)\overset{\Gamma=0}{=}\begin{pmatrix}   \frac{\omega+\epsilon\cosh{\left(2\sqrt{\epsilon^2-\omega^2}\cdot t\right)}}{\epsilon+\omega}&\frac{-\epsilon\sinh{\left(2\sqrt{\epsilon^2-\omega^2}\cdot t\right)}}{\sqrt{\epsilon^2-\omega^2}}\\\frac{-\epsilon\sinh{\left(2\sqrt{\epsilon^2-\omega^2}\cdot t\right)}}{\sqrt{\epsilon^2-\omega^2}}&\frac{-\omega+\epsilon\cosh{\left(2\sqrt{\epsilon^2-\omega^2}\cdot t\right)}}{\epsilon-\omega}
    \end{pmatrix}\,,
\end{equation}
while the mean number of photons is given by:
\begin{equation}
N(t)=\frac{1}{4}(\Sigma_{11}+\Sigma_{22})-\frac{1}{2}=\frac{\epsilon^2 \sinh^2\left(\sqrt{\epsilon^2-\omega^2}\cdot t\right)}{\epsilon^2-\omega^2}\,.
\end{equation}
Above formulas are valid for any $\epsilon$, while for $\epsilon<\omega$ it is worth to use equality $\sinh^2\left(\sqrt{\epsilon^2-\omega^2}\right)/(\epsilon^2-\omega^2)=\sin^2\left(\sqrt{\omega^2-\epsilon^2)}\right)/(\omega^2-\epsilon^2)$.

For $\epsilon>\omega$, the solution \eqref{eq:covgamma0} corresponds to the actual evolution of the physical system only for short times, as later, with indefinitely increasing number of photons, the quadratic term $\chi$ in \eqref{CQHam} becames non-negligible. Here, we focus on this short time. The number of photons in this case increases exponentially in time, and it is given by
\begin{equation}\label{meanNabove}
N_{\epsilon>\epsilon_c}(t) = \frac{\epsilon^2 \sinh^2\left(\sqrt{\epsilon^2-\omega^2}\cdot t\right)}{\epsilon^2-\omega^2} \sim  \frac{e^{2\sqrt{\epsilon^2-\omega^2} \cdot t}}{4}\,,
\end{equation}
where the asymptotic expansion holds for $t\gg(\sqrt{\epsilon^2-\omega^2})^{-1}$ and $\epsilon\gg\omega$.

For $\epsilon<\omega$, $\sqrt{\epsilon^2-\omega^2}$ becomes imaginary and we get:
\begin{equation}
N(t)=\frac{1}{4}(\Sigma_{11}+\Sigma_{22})-\frac{1}{2}=\frac{\epsilon^2 \sin^2\left(\sqrt{\omega^2-\epsilon^2}\cdot t\right)}{\omega^2-\epsilon^2}\,,
\end{equation}
so the number of photons oscillates periodically. However, while approaching the critical point, for $\epsilon\to \omega$, 
the period extends for any length of time, so,
for $t\ll (\sqrt{\omega^2-\epsilon^2})^{-1}$, we obtain:
\begin{equation}
N(t)\approx\omega^2t^2\,.
\end{equation}

\section{Quantum Fisher Information for CQS at zero temperature and the optimal measurement}\label{IV}
\label{IVb}
 We write the expression of the QFI only in the noiseless scenario. Indeed, for the dissipative scenario, its expression is very large, and we provide only an asymptotic analysis. Let us then consider separately the noiseless and the noisy scenarios.

\subsection{The noiseless case ($\Gamma=0$, $n_{B}=0$)}
In the noiseless case, the covariance matrix of the mode \(a(t)\) is given by \eqref{eq:covgamma0}. Inserting \(\bf\Sigma\) in \eqref{QFIGaussian} leads to
\begin{align}
I_{\rm cr}&=\frac{\epsilon^2\left[\epsilon^2(3+8\omega_0^2t^2)-4\omega_0^2(1+2\omega_0^2t^2)-4(\epsilon^2-\omega_0^2)\cosh{\left(2\sqrt{\epsilon^2-\omega_0^2}\cdot t\right)}+\epsilon^2\cosh{\left(4\sqrt{\epsilon^2-\omega_0^2}\cdot t\right)}\right]}{4(\epsilon^2-\omega_0^2)^3}+\nonumber\\\quad&-\frac{\epsilon^2\left[8\omega_0 t\sqrt{\epsilon^2-\omega_0^2}\sinh{\left(2\sqrt{\epsilon^2-\omega_0^2}\cdot t\right)}\right]}{4(\epsilon^2-\omega_0^2)^3}\,.
\end{align}

If we consider \(\epsilon<\epsilon_c\) (notice that in the noiseless case \(\epsilon_c=\omega_0\)), the asymptotic analysis reveals that
\begin{align}
I_{\rm cr}^{\epsilon<\epsilon_c}\sim \left[2N(t)+\frac{8N^2(t)}{9}\right]t^2\,, \quad\quad \epsilon\to \epsilon_c^{-}\,,
\end{align}
where $N(t)\simeq \omega^2_0 t^2$. With a bound $N_{\rm max}$ on the number of photons, the optimal choice is to set $\omega_0=\sqrt{N_{\rm max}}/T$ for total time $T$, which gives $I_{\rm cr}^{\epsilon<\epsilon_c}=O(N_{\rm max}^2T^2)$.

The situation changes for $\epsilon>\epsilon_c$. An asymptotic analysis  reveals that  
\begin{align}
I_{\rm cr}^{\epsilon>\epsilon_c}\sim \frac{4 N^2(t)}{\epsilon^2-\epsilon_c^2}\,, \quad\quad t\gg (\sqrt{\epsilon^2-\epsilon_c^2})^{-1}\,.
\end{align} 
Given the bound $N_{\rm max}$ on the number of photons \eqref{meanNabove}, we get $\epsilon \simeq \epsilon_c+\log^2(4N_{\rm max})/4T^2$ for total time $T$. Therefore, the QFI is $O(N_{\rm max}^2T^2/\log^2(N_{\rm max}))$, and the regime $\epsilon>\epsilon_c$ offers a worse performance than the case $\epsilon<\epsilon_c$.

\subsection{The dissipative case ($\Gamma>0$ , $n_{B}=0$)}
The QFI for the critical strategy can be computed using \eqref{QFIGaussian}. The expression is too long to be shown. However, we have performed asymptotic analysis for $\omega_0\leq \epsilon\leq \epsilon_c$ and $n_{B}=0$. Here, the QFI smoothly passes from $I_{\rm cr}\simeq \frac{2\omega_0^2}{(2\epsilon_c^2-\epsilon^2)\epsilon^2}N^2(t)$ for $t\simeq \lambda_+^{-1}$, to $I_{\rm cr}\simeq  \frac{8\omega_0^2}{(2\epsilon_c^2-\epsilon^2)\epsilon^2}N^2(t)$ for $t\simeq \lambda_-^{-1}$, until saturating to $I_{{\rm cr},0}\simeq \frac{8\omega_0^2}{(2\epsilon_c^2-\epsilon^2)\epsilon^2}N^2(\infty)$ at the steady state. $N(t)$ is a monotonic function of time, saturating at $N(\infty)=\frac{\epsilon^2}{2(\epsilon_c^2-\epsilon^2)}$, see Section~\ref{Ntcr}. The QFI rate is maximal when $\omega_0^2/(2\epsilon_c^2-\epsilon^2)\epsilon^2$ is maximized. By setting $\epsilon^2=z(\omega_0^2+\Gamma^2)$, for some $0<z<1$, one can easily see that $\omega_0=\Gamma$ maximizes the QFI. 

\begin{figure}[t!]
  \includegraphics[width=1
\textwidth]{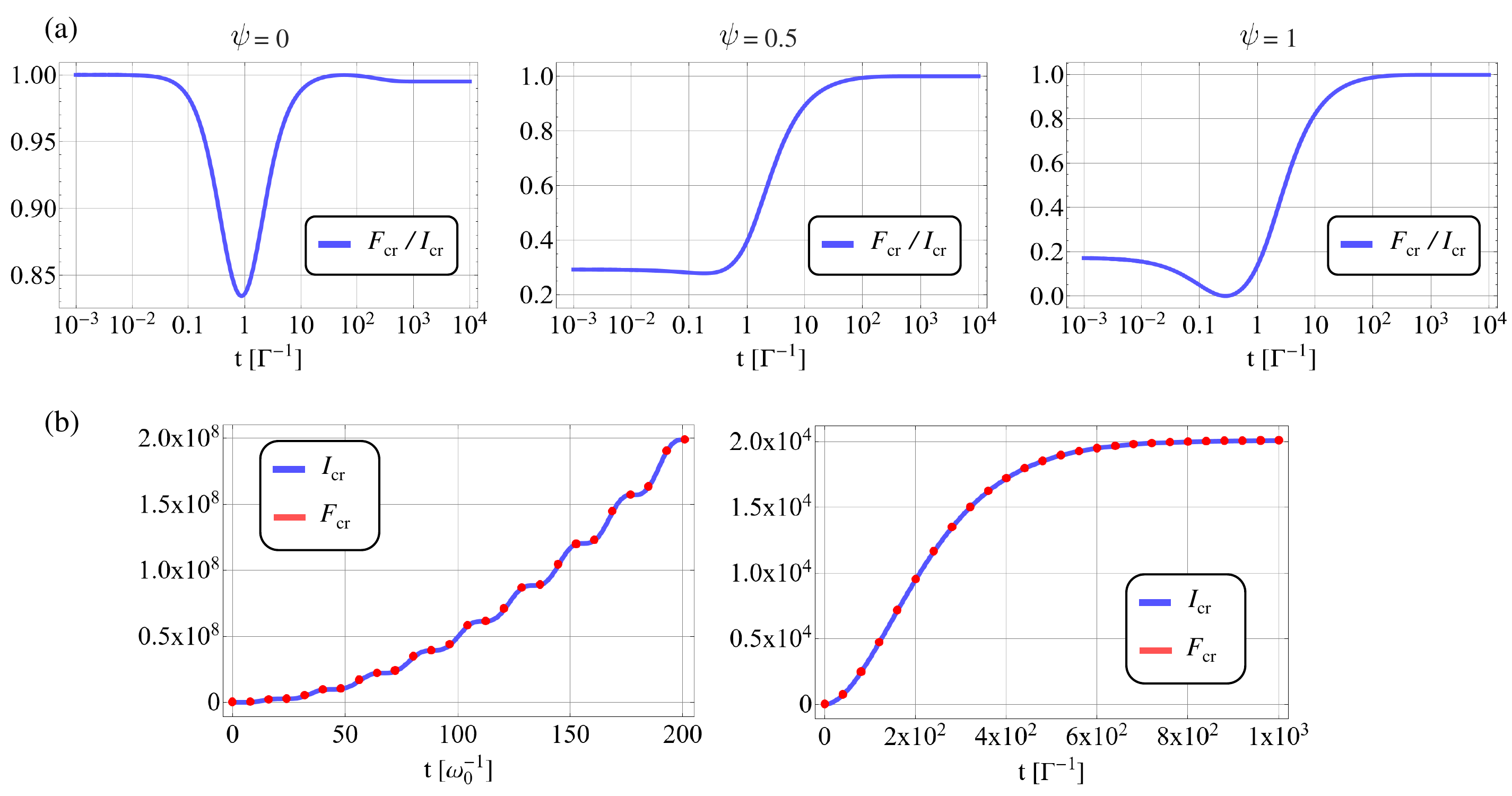}
\centering
\caption{\textbf{Comparison between homodyne FI and QFI for CQS.} \textbf{a)} We have plotted the ratio for different values of the quadrature angle \(\psi\), at zero temperature and with \(\omega_0=\Gamma\), \(\epsilon=\epsilon_{\rm opt}\), \(N_{\rm max}=100\), \(n_{B}=0\). \textbf{b)} We compare the QFI with the homodyne FI \textit{optimized} over the quadrature angle \(\psi\) in the noiseless case (\textit{left}, \(\epsilon=0.99\omega_0\)) and in the dissipative case (\textit{right}, \(\epsilon=\epsilon_{opt}\), \(N_{\rm max}=100\), \(n_B=0\)). The plots show that homodyne detection with an optimized angle is the optimal measurement.}
\label{Fig7}
\end{figure}

\subsection{Homodyne as optimal measurement}\label{Homodyne}

\subsubsection{Noiseless case}

For $\Gamma=0$ and $n_B=0$, the classical FI is given by (using Eq.~\eqref{FI}), 
\begin{align}
F^{\rm hom}_{\rm cr} \overset{\epsilon\to \epsilon_c}{\sim} \frac{2\omega^2 t^4\left[(3-2\omega_0^2t^2)\cos(2\psi)+2\omega_0 t(\omega_0 t +2\sin(2\psi))\right]^2}{9\left[4\omega_0^2t^2\sin^2(\psi)+2\omega_0 t\sin(2\psi)\right]^2}.
\end{align}
The FI optimized with respect to $\psi$ at each time virtually saturates the QFI, see Fig.~\ref{Fig7}.

\subsubsection{Steady state}
Before going to the analysis of the time evolution, let us briefly remind the results obtained for the steady-state \eqref{eq:covsteady} in \cite{DiCandia2023}. The steady state QFI is
\begin{equation}
    I_{\rm cr}=\frac{\epsilon^2}{2\epsilon_c^2-\epsilon^2}\left[\frac{1}{\epsilon^2_c-\epsilon^2}+\frac{2\omega_0^2}{(\epsilon_c^2-\epsilon^2)^2}\right]\overset{\epsilon\to \epsilon_c}{\sim}\frac{8\omega_0^2}{\epsilon_c^2}N^2(\infty)\,,
\end{equation}
and it is asymptotically saturated (while approaching the critical point) by the homodyne detection for arbitrary direction. In this case, the classical FI is given by (using~\eqref{FI})
\begin{equation}
    F^{\rm hom}_{\rm cr}=
    \frac{\epsilon^2[(\Gamma^2-\omega^2-\epsilon^2)\cos(2\psi)+2\omega\epsilon+2\omega\Gamma 
    \sin(2\psi)]^2}{2(\epsilon_c^2-\epsilon^2)^2[\epsilon_c^2-\epsilon(\omega\cos(2\psi)-\Gamma\sin(2\psi))]^2}    
    \overset{\epsilon\to \epsilon_c}{\sim}\frac{8\omega_0^2}{\epsilon_c^2}N^2(\infty)\,.
\end{equation}
Looking at the covariance matrix \eqref{eq:covsteady}, one should notice that changes of $\omega=\omega_0+\delta \omega$ result in the rotation of the covariance matrix, as well as in an increasing of the variances values. However, these effects are irrelevant compared to the very rapid change of the average number of photons for \(\epsilon\to\epsilon_c\), being \(N\propto\frac{1}{\epsilon_c^2-\epsilon^2}\).

It is therefore reasonable to ask whether almost all the information about the value of $\delta\omega$ can be obtained from the average number of photons. To answer this question, we analyze the signal-to-noise ratio for the mean number of photons (the quantity corresponding to the classical FI, but without the optimization over the estimator), which shows:

\begin{equation}
\frac{|\partial_\omega N|^2}{\Delta^2 N}=\frac{4\epsilon^2 \omega_0^2}{(3\epsilon_c^2-\epsilon^2)(\epsilon_c^2-\epsilon^2)^2} = \frac{16\omega_0^2}{\epsilon^2(3\epsilon_c^2-\epsilon^2)}N^2(\infty) \overset{\epsilon\to \epsilon_c}{\sim}  \frac{8\omega_0^2}{\epsilon_c^4}N^2(\infty)\,.
\end{equation}
This means that, close to the critical point, photon-counting is another example of optimal measurement. However, from a practical point of view, it is often easier to perform homodyne detection under realistic circumstances. Therefore, in further discussions, we focus only on homodyne detection.

\subsubsection{Time evolution}
 In Fig.~\ref{Fig7}, we compare the FI for homodyne detection (calculated using \eqref{FI}) and the QFI at zero temperature, for different values of \(\psi\). In addition, we show that homodyne detection, optimized with respect to \(\psi\) at each time, essentially saturates the QFI.

\section{The passive system}
\label{sec:passive}
\subsection{The model and the evolution}

In the passive case, we consider the Hamiltonian in \eqref{HSchro} with $\epsilon=0$, $\chi=0$ and dissipations as in \eqref{noise_suppl}. We initialize the system to a state with $N_{\rm max}$ number of photons, i.e., $\rho=D(\alpha)S(r)\rho_B S^\dag (r)D^\dag (\alpha)$. Here, $D(\alpha)=e^{\alpha a^\dag -\alpha^* a}$  with $\alpha=|\alpha| e^{i\phi}$, is a displacement operator, and $S(r)=e^{\frac{1}{2}(r a^{\dag 2}-r^*a^2)}$ is a squeezing operator. The state $\rho_B=\sum_{k=0}^\infty \frac{n_{B}^k}{(1+n_{B})^{1+k}}|k\rangle\langle k|$ is a thermal state with $n_{B}$ average number of photons. Without loss of generality, we choose $r$ to be real and positive. The number of photons is given by $N_{\rm max}=|\alpha|^2+\sinh^2(r)(2n_{B}+1)+n_{B}$, which will act as a constraint in the computation of the QFI. 

The evolution of such a passive system can be easily written in the frame rotating with $\omega_r$ as 
\begin{align}
a(t) = e^{-\Gamma t-i\delta \omega t}a(0) +\sqrt{1-e^{-2\Gamma t}} b\,,
\end{align}
where $b$ is a thermal mode with $n_{B}$ photons.  The first-moment vector and the covariance matrix are given by Eqs.~\eqref{fmv}-\eqref{covmatr}, by substituting
\begin{align}
\langle a(t)\rangle &= e^{-\Gamma t-i\delta \omega t}\alpha\,,\\
\langle a^2(t)\rangle &= e^{-2\Gamma t}e^{-2i\delta \omega t}\left[
 \frac{1}{2}\sinh{(2r)}(2n_{B}+1)+\alpha^2\right]\,,\\
\langle a^\dag (t)a(t)\rangle &= |\alpha|^2+\sinh^2(r)(2n_{B}+1)+n_{B}\label{fmv passive strategy}\,.
\end{align}

\subsection{QFI for PQS}\label{IVa}
Since the mode $a(t)$ is Gaussian, we can analytically compute the QFI using \eqref{QFIGaussian}. By setting the derivative with respect to $\phi$ to zero, we realize that $\phi=0$, i.e., a real $\alpha$, is the optimal choice. The formula for generic $n_{B}$ is large to be shown. Let us first consider $n_{B}=0$. We have that
\begin{align}\label{QFIpas0}
I_{\rm pas}= \left[\frac{4\alpha^2}{e^{-2r}+e^{2\Gamma t}-1}+\frac{e^{-2r}(e^{4r}-1)^2}{2e^{2r+4\Gamma t}+(e^{2r}-1)^2(e^{2\Gamma t}-1)}\right]t^2\,.
\end{align}
Let us optimize the QFI in the noiseless and noisy cases separately. 

\subsubsection{The noiseless scenario ($\Gamma=0$, $n_{B}=0$)}
In this case, we get  
\begin{align}
I_{\rm pas}  = \left[4\alpha^2 e^{2r}+2\sinh^2(2r)\right]t^2\,.
\end{align}
This can be maximized with the constraint $N_{\rm max}=\alpha^2+\sinh^2(r)$, obtaining that the squeezed-vacuum state ($\alpha=0$) is optimal. We obtain then
\begin{align}\label{optQFIless}
I_{\rm pas} =  8N_{\rm max}(1+N_{\rm max})t^2\,.
\end{align}
This holds also for the noisy case, as long as $N_{\rm max}\Gamma t\ll1$. In Section~\ref{HomPQS}, we will see that in order to saturate this QFI, non-linear detection is needed.

\subsubsection{The dissipative scenario ($\Gamma>0$, $n_{B}=0$)}\label{noisyQFI}

In the noisy case, we first approximate the QFI \eqref{QFIpas0} with $e^{2r}\gg1$: 
\begin{align}
I_{\rm pas}\simeq \left[\frac{4\alpha^2}{e^{-2r}+e^{2\Gamma t}-1}+\frac{e^{4r}}{2e^{4\Gamma t}+e^{2r}(e^{2\Gamma t}-1)}\right]t^2\,.
\end{align}
In the $e^{2r}\gg e^{4\Gamma t}/(e^{2\Gamma t}-1)$ regime, we have that 
\begin{align}\label{optQFI0}
I_{\rm pas}\simeq \frac{\left(4\alpha^2+e^{2r}\right)t^2}{e^{2\Gamma t}-1} \simeq \frac{4N_{\rm max}t^2}{e^{2\Gamma t}-1}\,.
\end{align}
This means that the QFI is maximized by any state satisfying the condition $e^{2r}\gg e^{4\Gamma t}/(e^{2\Gamma t}-1)$. In practice, the QFI maximum is reached at $t\simeq 0.8/\Gamma$, meaning that the conditions correspond to a relevant regime.

\subsection{Homodyne FI for PQS}\label{HomPQS}

\begin{figure}[t!]
  \includegraphics[width=1
\textwidth]{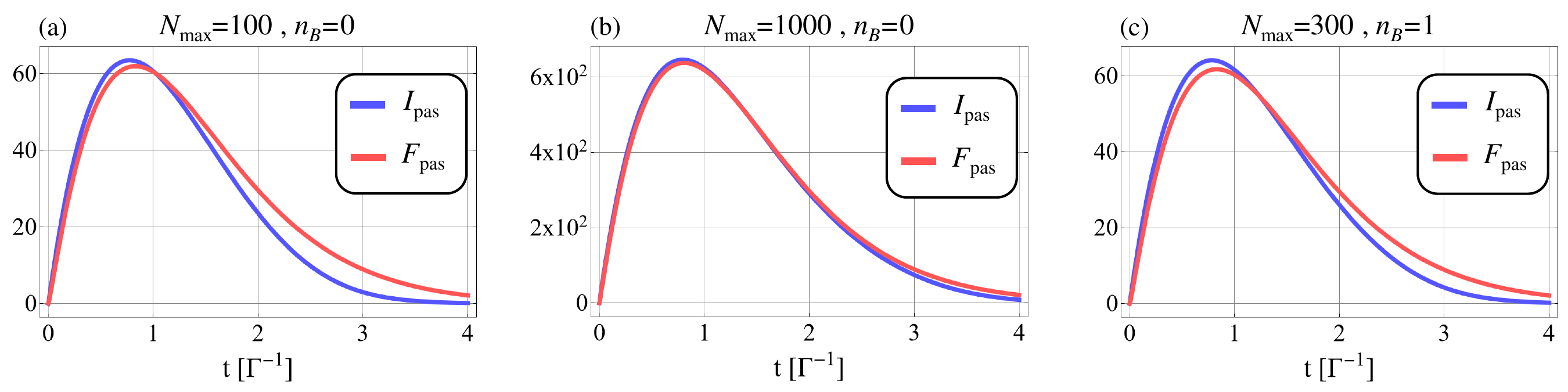}
\centering
\caption{\textbf{Homodyne FI vs QFI.} {\bf(a)} Comparison between \(I_{\rm pas}\) with $(\alpha=0, r\simeq 3$) and \(F_{\rm pas}\) optimized over both $\alpha,r$ for \(N_{\rm max}=100\) and $n_{B}=0$. Here, $F_{\rm pas}$ surpasses $I_{\rm pas}$ at longer time, as the condition $e^{2r}\gg e^{4\Gamma t}/(e^{2\Gamma t}-1)$ is clearly not satisfied already for $t\gtrsim 2\Gamma^{-1}$. However, for $t\lesssim \Gamma^{-1}$, the condition on $r$ is satisfied, and $I_{\rm pas}$ is optimal. {\bf(b)} Comparison between \(I_{\rm pas}\) with $(\alpha=0, r\simeq 4.15$) and the optimized \(F_{\rm pas}\), for \(N_{\rm max}=1000\) and $n_{B}=0$. Homodyne saturates the optimal QFI already for $N_{\rm max}=10^3$. {\bf(c)} Comparison between \(I_{\rm pas}\) with $(\alpha=0, r\simeq 3$) and the optimized \(F_{\rm pas}\), for \(N_{\rm max}=300\) and \(n_{B}=1\). The plot is very similar to the zero-temperature case with $100$ photons (corresponding to $N_{\rm max}/(1+2n_{B})$), see {\bf (a)}.}
\label{PasVSHomo}
\end{figure}

In this case, we consider a protocol where $\alpha$ and $r$ are real, i.e., we displace along the $x$ quadrature and squeeze along the $p$ quadrature, and we measure $p$, i.e., we choose the quadrature angle $\psi=\pi/2$. We obtain 
\begin{align}\label{FIpas}
F_{\rm pas} =\frac{4\alpha^2t^2}{(1+2n_{B})(e^{-2r}+e^{2\Gamma t}-1)}\,,
\end{align}
which can be maximized with the constraint $N_{\rm max}=\alpha^2+\sinh^2(r)(1+n_{B})+n_{B}$. 

In the noiseless scenario, i.e., for $\Gamma=0$ and $n_{B}=0$, we get $F_{\rm pas}=4N_{\rm max}(1+N_{\rm max})t^2$, which is a constant worse then the optimal QFI in \eqref{optQFIless}. In this case, homodyne detection does not saturate the QFI.

For the noisy scenario, one can optimize with respect to $r$ under the constraint on $N_{\rm max}$. For instance, for $n_{B}=0$, we have the optimal squeezing 
\begin{equation}\label{opt_r}
    r= \log\left({\sqrt{\frac{1}{2}\left(\coth{(\Gamma t)}-1\right)\left(\sqrt{e^{4\Gamma t}+4N_{\rm max}(e^{2\Gamma t}-1)}-1\right)}}\right)\,. 
\end{equation}
The optimal FI in this regime is
\begin{align}\label{QFIpassiveoptimal}
F_{\rm pas}&\simeq \frac{8N_{\rm max}(1+N_{\rm max})t^2}{e^{2\Gamma t}(1+2N_{\rm max})-2N_{\rm max}+\sqrt{e^{4\Gamma t}+4N_{\rm max}(e^{2\Gamma t}-1)}} \\
\quad&\sim \frac{4N_{\rm max}t^2}{e^{2\Gamma t}-1}\,, \label{asymp2}
\end{align}
where the asymptotic expansion is for large $N_{\rm max}$. Notice that this strategy is asymptotically optimal, as it matches the optimal QFI in \eqref{optQFI0}. 

For generic $n_{B}$, it is still possible to find a close, but lengthy expression for the optimal squeezing. However, we shall notice that by choosing $e^{2r}\gg (e^{2\Gamma t}-1)^{-1}$, we get $F_{\rm pas}\simeq 4\alpha^2t^2/(1+2n_{B})(e^{2\Gamma t}-1)$. If we choose $\alpha^2\simeq N_{\rm max}$, we get that 
\begin{align}
F_{\rm pas}\simeq \frac{4N_{\rm max}t^2}{(1+2n_{B})(e^{2\Gamma t}-1)}\,,
\end{align}
which matches the optimal QFI in \eqref{optQFIpasNT}.

Fig. \ref{PasVSHomo} compares the optimal homodyne strategy with the squeezed-vacuum strategy, which is optimal both for very short times and when the condition $e^{2r}\gg e^{4\Gamma t}/(e^{2\Gamma t}-1)$ is satisfied.  Homodyne reaches essentially the same precision with a small time lag. Nevertheless, this lag is also responsible for the advantage of the squeezed-vacuum state in Fig.~3 of the main text (blue line). For large $N_{\rm max}$, we have seen in Eqs.~\eqref{optQFI0}-\eqref{asymp2} that the two strategies are equivalent. This is visible already for $N_{\rm max}=10^{3}$, as shown in Fig.~\ref{PasVSHomo}.
The result is unchanged when considering finite temperatures, aside from a dividing factor $(1+2n_{B})$ for both optimal and homodyne based strategies.

\section{Optimal measurement time}
\label{optimal}

\subsection{Single-shot case}

Here, we allow the system to evolve for an arbitrary time, and a single measurement is performed at the end. This corresponds to saying that time is not seen as a resource and the only resource is the total average number of photons in the system \(N_{\rm max}\). Nevertheless, there is an optimal measurement time. 

In the noiseless case, we have seen in the main text that the protocol can be carried out coherently, as the QFI grows as $\propto T^2$.

In the dissipative case at zero temperature, Fig.~2 of the main text shows very well the maximal points. As for PQS, we have that $I_{\rm pas}\sim 4N_{\rm max}t^2/(e^{2\Gamma t}-1)$, see \eqref{optQFI0}. This is maximal for $t\simeq 0.8\cdot\Gamma ^{-1}$, which gives us the optimal QFI $I_{\rm pas}^{\rm max}\simeq 0.65 N_{\rm max}/\Gamma^2$ for large $N_{\rm max}$. For the CQS, we have shown that the QFI rate is maximized at the steady state, therefore theoretically the optimal time is $t\to\infty$, but in practice, QFI does not change significantly after $t\simeq \lambda_-^{-1}$. The optimal QFI is achieved for $\omega_0=\Gamma$, and its value is $2N_{\rm max}^2/\Gamma^2$.
At arbitrary temperatures, time scales are unchanged, and we have that the QFI changes roughly by substituting $N_{\rm max}/(1+2n_{B})$ to $N_{\rm max}$.

\subsection{Multiple repetition case}
\label{sec:multiple}

We consider now both the total protocol time \(T\) and the number of photons \(N_{\rm max}\) as resources. In this case, we perform \(M\) measurements at times \(t=T/M\). Thus, the total QFI is $M\cdot I(t)=I(t)T/t$, where \(I(t)\) is the single shot QFI. The optimal precision will be achieved for the time maximizing $I(t)/t$. The same holds for the total FI $M\cdot F(t)$, where the optimal precision will be achieved for the time maximizing $F(t)/t$.

In the noiseless scenario, this optimization is trivially solved as $t=T$, as the QFI grows as $\propto T^2$. 

In the dissipative case at zero temperature, PQS asymptotically saturates the fundamental bound in Eq.5 of the main text, namely $I^{\rm total}_{\rm pas}\leq 2N_{\rm max}T/\Gamma$. This may be obtained in both cases with pure squeezing, as well as squeezing+displacement followed by homodyne detection.

Indeed, for pure squeezing, from \eqref{QFIpas0}, we have
\begin{equation}
    \frac{I_{\rm pas}}{t}=\frac{e^{-2r}(e^{4r}-1)^2t}{2e^{2r}e^{4\Gamma t}+(e^{2r}-1)^2(e^{2\Gamma t}-1)}\overset{e^{2r}\gg 1}{\approx} \frac{e^{2r}te^{-4\Gamma t}}{2e^{-2r}+(1-e^{-2\Gamma t})}\overset{\Gamma t\ll 1}{\approx}\frac{1}{\Gamma}\frac{e^{2r}\Gamma t}{2e^{-2r}+2\Gamma t}\overset{\Gamma t\gg e^{-2r}}{\approx} \frac{1}{\Gamma}\frac{e^{2r}}{2}\approx \frac{2N_{\max}}{\Gamma}\,,
\end{equation}
so the bound is saturated for $N_{\max}\gg1$, for single-repetition time satifying $1/(N_{\max}\Gamma)\ll t\ll 1/\Gamma$.

For homodyne with $N_{\max}=\alpha^2+\sinh^2r$, from \eqref{FIpas}, we have
\begin{align}
\frac{F_{\rm pas}}{t} =\frac{4\alpha^2t}{(e^{-2r}+e^{2\Gamma t}-1)}\overset{\Gamma t\ll 1}{\approx}\frac{1}{\Gamma} \frac{4\alpha^2\Gamma t}{e^{-2r}+2\Gamma t}\overset{\Gamma t\gg e^{-2r}}{\approx}\frac{2\alpha^2}{\Gamma}\overset{\alpha^2\gg \sinh^2(r)}{\approx}\frac{2N_{\max}}{\Gamma}\,,
\end{align}
so the bound is saturated for $N_{\max}\gg 1$, for single-repetition time satisfying $e^{-2r}/\Gamma\ll t\ll 1/\Gamma$, where the number of photons due to squeezing is small compared to the number of photons due to displacement, i.e., $\sinh^2(r)\ll \alpha^2.$

The CQS optimal time for the single-shot and multiple repetition case is very similar, see Fig.~2 and Fig.~3 in the main text.

For higher temperatures, time scales are the same, so nothing fundamentally changes.

\section{Fundamental bound for QFI for the estimation of the frequency of the cavity coupled to the thermal bath}
\label{sm:fundamental}

Here, we derive the fundamental bound to the precision obtainable for the estimation of the frequency of the cavity coupled to the thermal bath using total time $T$, with restriction of the average number of photons. We will use the theorem from \cite{demkowicz2017adaptive}. See also~\cite[App. E]{kurdzialek2022using} for an extension to the case where the parameter is encoded in Lindblad operators and~\cite{wan2022bounds} for an alternative derivation.

Let us first recall the result of \cite{demkowicz2017adaptive} we want to use. For a general time evolution of a quantum state \(\rho\), described by the Lindblad equation:
\begin{equation}
\frac{d\rho}{dt}=-i\omega[H,\rho]+\sum_{j=1}^JL_j\rho L_j^\dagger -\frac{1}{2}\rho L_j^\dagger L-\frac{1}{2}L_j^\dagger L\rho\,,
\end{equation}
we define the following operators:
\begin{align}
    \beta^{(1)} &= \dot{H} + h_{00}^{(1)} \openone + \vect{L}^\dag \vect{h}^{(\frac{1}{2})} + \vect{h}^{(\frac{1}{2})\dag} \vect{L} + \vect{L}^\dag \mathfrak{h}^{(0)} \vect{L}\,,\\\quad
    \alpha^{(1)} &= \left[ \vect{h}^{(\frac{1}{2})} \openone  + \mathfrak{h}^{(0)} \vect{L} \right]^\dag \left[\vect{h}^{(\frac{1}{2})} \openone  + \mathfrak{h}^{(0)} \vect{L}    \right]\,,
\end{align}
where $\vect{L}$ is the vector of Lindblad operators, $h_{00}^{(1)}$ is a scalar, $\vect{h}^{(\frac{1}{2})}$ is a vector of length $J$ and $\mathfrak{h}^{(0)}$ is a  $J\times J$ matrix mixing Lindblad operators with each others. 

Then, for any adaptive strategy involving entanglement with arbitrary large ancillas and acting with arbitrary unitaries during evolution, the QFI for the estimation of the parameter $\omega$, after an evolution of time $T$, is bounded by (Eq. (18) of \cite{demkowicz2017adaptive}):
\begin{equation}\label{boundh}
   I\leq 4\int_0^T \min_{\{h_{00}^{(1)},\vect{h}^{(\frac{1}{2})},\mathfrak{h}^{(0)}\}} \langle \alpha^{(1)}\rangle_t dt\,,\quad \textrm{subject to}\quad \beta^{(1)}=0\,.
\end{equation}
Note that such a general scheme includes dividing the total time $T$ into smaller parts, performing measurements in each part, and eventually updating the protocol based on these results. Note also that a minimization over $\{h_{00}^{(1)},\vect{h}^{(\frac{1}{2})},\mathfrak{h}^{(0)}\}$ gives the tightest bound, but \eqref{boundh} is valid for any choice of $\{h_{00}^{(1)},\vect{h}^{(\frac{1}{2})},\mathfrak{h}^{(0)}\}$.

Let us now go to the system discussed in this paper. To apply the above theorem, it is important to distinguish what is an unchangeable part of the evolution of our system and what is an additional, tunable part, connected with a peculiar strategy. In our case, the first one is:
\begin{equation}
\label{eq:lind}
\frac{d\rho}{dt}=-i\omega[a^\dagger a,\rho]+\Gamma (1+n_{B})\left(2a\rho a^\dag -\{a^\dag a,\rho\}\right)+\Gamma n_{B}\left(2a^\dag\rho a-\{aa^\dag,\rho\}\right)\,,
\end{equation}
while the second one is the  unitary squeezing $\tfrac{\epsilon}{2}(a^2+a^{\dag 2})$. More precisely, referring to Fig.~1 from \cite{demkowicz2017adaptive}, the gate $\mathcal E_t^\omega$ corresponds to integrating \eqref{eq:lind} over $t$, while the unitary control $U$ corresponds to integrating over $t$ the expression $\frac{d\rho}{dt}=-i[\tfrac{\epsilon}{2}(a^2+a^{\dag 2}),\rho]$. Note that even if these two operations do not commute for finite $t$, for $t\to 0$ applying them alternately $T/t$ times becomes equivalent to evolving $\frac{d\rho}{dt}=-i[\omega a^\dagger a+\tfrac{\epsilon}{2}(a^2+a^{\dag 2}),\rho]+\mathcal L[\rho]$.

Therefore, we have two Lindblad operators, namely $L_1=\sqrt{2\Gamma(1+n_{B})}a$, $L_2=\sqrt{2\Gamma n_{B}}a^\dagger$ and $\dot H=a^\dagger a$. Putting $\vect{h}^{(\frac{1}{2})}=0$, $\mathfrak{h}^{(0)}_{12}=\mathfrak{h}^{(0)}_{21}=0$, we get:
\begin{equation}
   \beta^{(1)}=a^\dagger a+h_{00}^{(1)} \openone +    
   2\Gamma (n_B+1)a^\dag a\mathfrak{h}^{(0)}_{11}+
   2\Gamma n_B a a^\dag \mathfrak{h}^{(0)}_{22}\,,
\end{equation}
which is zero under the conditions:
\begin{equation}
\label{eq:beta0}
   \begin{split}
       1+2\Gamma (n_B+1)\mathfrak{h}^{(0)}_{11}+2\Gamma n_B  \mathfrak{h}^{(0)}_{22}=0\,,\\
       h_{00}^{(1)}+2\Gamma n_B  \mathfrak{h}^{(0)}_{22}=0\,.
   \end{split}
\end{equation}
We have also:
\begin{equation}
   \alpha^{(1)}=2\Gamma (n_B+1)a^\dag a\left(\mathfrak{h}^{(0)}_{11}\right)^2+2\Gamma n_B a a^\dag \left(\mathfrak{h}^{(0)}_{22}\right)^2\,.
\end{equation}
Setting $\langle a^\dagger a\rangle_t=N(t)$, after a direct minimization with the constraints in \eqref{eq:beta0}, we obtain:
\begin{equation}
   \langle\alpha^{(1)}\rangle_t=\frac{N(t)}{2\Gamma(1+2n_B-\frac{n_B}{N(t)+1})}\,,
\end{equation}
and therefore:
\begin{equation}
   I\leq \int_0^T \frac{2N(t)}{\Gamma(1+2n_B-\frac{n_B}{N(t)+1})} dt\leq \frac{2N_{\max}T}{\Gamma(1+2n_B-\frac{n_B}{N_{\max}+1})}   
   \,,
\end{equation}
where in the last step we used the fact, that the function under the integral is strictly increasing with $N(t)$. As discussed in Section~\ref{optimal}, this bound is saturable by the passive strategy in the limit of a large number of photons.

\section{Time to charge and discharge a linear resonator}\label{timechargedischarge}
Assume we have a linear resonator in a vacuum state, and we want to charge the cavity with a state prepared outside the cavity. This situation is modeled by the Langevin equation, that, in the frame rotating with the resonator frequency, is given by 
\begin{equation}
    \Dot{a}(t)=\Gamma a(t)-\sqrt{2\Gamma}a_{\rm in}(t)\,,
\end{equation}
where $a(0)$ is in the vacuum state and $a_{\rm in}(\tau)$ is the input signal of the resonator. Here, we have considered the favorable case where all the cavity losses are due to the interaction with the preparation line.
This equation is solved as 
\begin{align}
a(t)&=e^{-\Gamma t} a(0)+\sqrt{2\Gamma}\int_0^td\tau\, e^{-\Gamma( t-\tau)}a_{\rm in}(\tau)= e^{-\Gamma t}~a(0)+\sqrt{(1-e^{-2\Gamma t})}~a_P,
\end{align}
where $a_{P}=\sqrt{\frac{2\Gamma}{1-e^{-2\Gamma t}}}~\int_0^td\tau e^{-\Gamma( t-\tau)}a_{\rm in}(\tau)$ is the integrated input mode. Assuming that $\langle a_{P}^\dag a_{P}\rangle =N_{\rm max}$, and we want to prepare the state up to an error $\epsilon$ in the number of photons, i.e., we want to achieve $\langle a^\dag(t) a(t)\rangle = N_{\rm max}-\xi$, this takes a time given by the solution of
\begin{align}
N_{\rm max}-\xi = (1-e^{-2\Gamma t})N_{\rm max},
\end{align}
i.e., $t=\log(N_{\rm max}/\xi)/2\Gamma$. The same time is needed to discharge the resonator in a state with $N_{\rm max}$ photons.

\section{Quantum Fisher Information at finite temperature}\label{noiseQFI}

\begin{figure}[t!]
  \includegraphics[width=1
\textwidth]{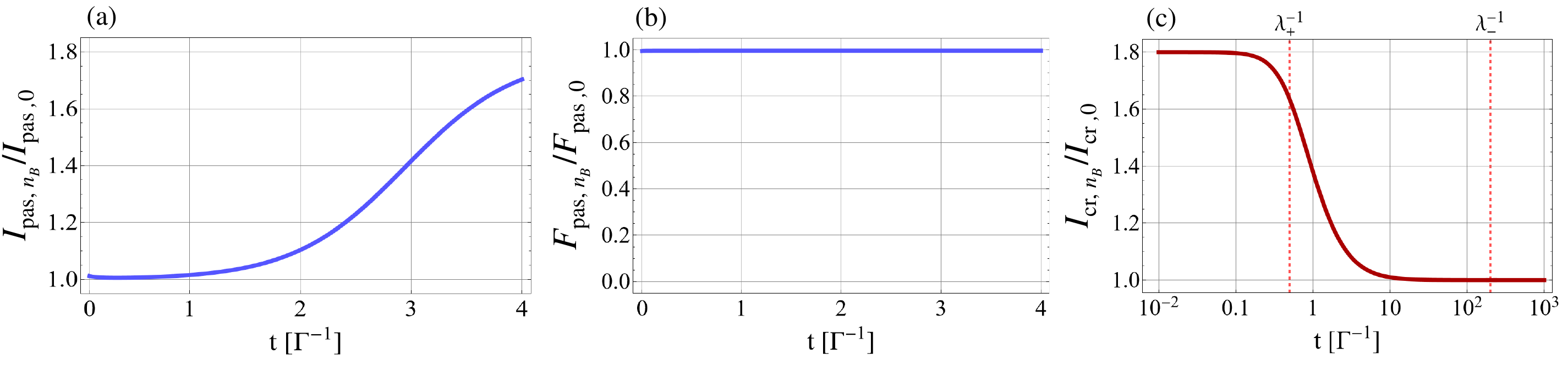}
\centering
\caption{\textbf{QFI at arbitrary temperature.} {\bf (a)} Ratio between the QFIs  \(I_{{\rm pas},n_B}\) and \(I_{{\rm pas},0}\), considering squeezed-vacuum as initial state. 
Here, and in the following, we denote with the subscript \(n_B\) the QFI at finite temperature and with the subscript \(0\) the QFI at zero temperature. The parameters \(\alpha\) and \(r\) are the same for both QFIs.
Notice that $I_{{\rm pas},0}$ corresponds to a strategy with a mean number of photons smaller by a factor $(1+2n_{B})$ with respect to the strategy for \(I_{{\rm pas},n_B}\). {\bf (b)} Ratio between the homodyne FIs \(F_{{\rm pas},n_B}\) and \(F_{{\rm pas},0}\). {\bf (c)} Ratio between \(I_{{\rm cr},n_B}\) and \(I_{{\rm cr},0}\) at the optimal point $\epsilon=\epsilon_{\rm opt}\simeq 0.9975$. The optimal point is the same for both the finite- and zero-temperature cases. To obtain the various plots, we have set: \(N_{\rm max}=300,\,n_{B}=1,\,\omega_0=\Gamma,\,\epsilon=0.9975\epsilon_c\). In {\bf (a)} we have set $\alpha=0$ and $r\simeq 3$, while in {\bf (b)} we have set the parameters optimizing the homodyne FI, see \eqref{opt_r}.}
\label{FigNoisy}
\end{figure}

\subsection{CQS}
We consider the case where all the parameters of the system, namely $\omega_0$, $\Gamma$ and $\epsilon$ are the same, and we distinguish two situations: one with a zero-temperature bath $n_B=0$ (we denote the corresponding QFI as $I_{{\rm cr},0}$) and one with a finite temperature bath $n_B$ ($I_{{\rm cr},n_B}$). In both cases, the protocol starts with the proper thermal state (which is the vacuum in the first case). Notice that the finite temperature does not modify the critical point, whose expression depends only on \(\Gamma\) and \(\omega_0\), and is \(\epsilon_c=\sqrt{\omega_0^2+\Gamma^2}\) even for \(n_B>0\).

At arbitrary temperature, the QFI $I_{{\rm cr},n_B}\simeq 2I_{{\rm cr},0}$ for $t\simeq \lambda_+^{-1}$, and $I_{{\rm cr},n_B}$ rapidly approaches $I_{{\rm cr},0}$ for $t\gtrsim \lambda_+^{-1}$, see Fig.~\ref{FigNoisy}. The mean number of photons is $N_{n_{B}}(t)\simeq (1+2n_{B})N_{n_B=0}(t)+n_{B}\simeq (1+2n_{B})N_{n_B=0}(t)$ for large enough $N_{n_B=0}(t)$, see Section~\ref{NumnT}. This means that $I_{{\rm cr},n_B}$ is roughly equal to $I_{{\rm cr},0}$, but it is obtained for $(1+2n_{B})$ times bigger number of photons, i.e., $N_{\max}'=(1+2n_B)N_{\max}$.

\subsection{PQS}

In the PQS case, we can do the same type of analysis as in Section~\ref{noisyQFI}. We consider $e^{2r}\gg e^{4\Gamma t}/(e^{2\Gamma t}-1)$, obtaining 
\begin{align}\label{optQFIpasNT}
I_{\rm pas}\simeq \frac{4N_{\rm max}t^2}{(1+2n_{B})(e^{2\Gamma t}-1)}\,,
\end{align}
where $N_{\rm max}=|\alpha|^2+(1+2n_{B})\sinh^2(r)+n_{B}$. 
As discussed in Section \ref{noisyQFI}, this asymptotic scaling may be obtained for a broad choice of the system parameters, including both situations where $\sinh^2(r)\gg |\alpha|^2$ or $|\alpha|^2\gg \{\sinh^2(r),n_B\}$. Detailed analysis shows that, for finite $N_{\max}$, the FI is slightly better if we consider pure squeezing. To compare it with the above discussion about CQS, we need to point out some points.

In CQS, for the same system parameters, but at finite temperature, the number of photons was roughly rescaled by a factor $(1+2n_B)$. In PQS, if we fix the parameters $\alpha$ and \(r\), an analogous situation holds only if 
$\sinh^2(r)\gg |\alpha|^2$. Then, the QFI is the same as in the noiseless case, but with the number of photons $(1+2n_B)$ times bigger. Indeed, in this case, PQS behaves in the same way as CQS. However, if $|\alpha|^2\gg \{\sinh^2(r),n_B\}$, a finite temperature does not affect significantly the number of photons (see \eqref{fmv passive strategy}). 
Therefore, obtaining the same value of the QFI would require changing the values of the parameters $\alpha$ and \(r\).\\
 In both cases, to obtain the same precision, one needs to properly increase the number of maximal allowed average number of photons $N_{\max}'=(1+2n_B)N_{\max}$, as in CQS.

\end{document}